\documentclass[a4paper,11pt]{article}
\pdfoutput=1 % if your are submitting a pdflatex (i.e. if you have
\usepackage{jheppub} % for details on the use of the package, please
\usepackage[T1]{fontenc} % if needed
\usepackage{diagbox}
\usepackage{multirow}
\usepackage{slashed}
\usepackage{fix-cm}
\usepackage{graphicx}
\usepackage{rotating}
\usepackage{longtable,pdflscape}
\usepackage{stackengine}
\usepackage[export]{adjustbox}

\allowdisplaybreaks[1]

\def\be{\begin{equation}}
\def\ee{\end{equation}}
\def\beq{\begin{eqnarray}}
\def\eeq{\end{eqnarray}}

\newcommand{\gam}{\gamma}

\newcommand{\msbar}{\overline{\text{MS}}}
\def\nn{\nonumber}

\def\mH{\mathcal{H}}
\def\mC{\mathcal{C}}
\def\mQ{\mathcal{Q}}

\def\gev{{\rm GeV}}
\def\mev{{\rm MeV}}

\usepackage{tikz}
\usetikzlibrary{calc}%%%用于坐标计算，否则不能运行($(c1)-(0.5,0)$)--($(c1)+(0.5,0)$)
\usetikzlibrary{trees}
\usetikzlibrary{decorations.pathmorphing}
\usetikzlibrary{decorations.markings}
\usetikzlibrary{arrows.meta}%%箭头类型设置“>=Stealth”必须调用这个宏包，否则报错
\tikzset{
    vector/.style={decorate, decoration={snake}, draw},
	provector/.style={decorate, decoration={snake,amplitude=2.5pt}, draw},
	antivector/.style={decorate, decoration={snake,amplitude=-2.5pt}, draw},
    fermion/.style={draw=black, postaction={decorate},
        decoration={markings,mark=at position .5 with {\arrow[draw=black]{>}}}},
    fermionbar/.style={draw=black, postaction={decorate},
        decoration={markings,mark=at position .5 with {\arrow[draw=black]{<}}}},
    fermionnoarrow/.style={draw=black},
    gluon/.style={decorate, draw=black,
        decoration={coil,amplitude=4pt, segment length=5pt}},
    scalar/.style={dashed,draw=black, postaction={decorate},
        decoration={markings,mark=at position .5 with {\arrow[draw=black]{>}}}},
    scalarbar/.style={dashed,draw=black, postaction={decorate},
        decoration={markings,mark=at position .5 with {\arrow[draw=black]{<}}}},
    scalarnoarrow/.style={dashed,draw=black},
    electron/.style={draw=black, postaction={decorate},
        decoration={markings,mark=at position .5 with {\arrow[draw=black]{>}}}},
	bigvector/.style={decorate, decoration={snake,amplitude=4pt}, draw},
}
\title{\bf \boldmath Revisiting $B_{c}^-\to J/\psi (\eta_c) L^-$ decays within the SM and beyond in QCD factorization}

\author[a,d]{Wei-Jun Deng,}
\author[b]{Fang-Min Cai,}
\author[c,e,1]{Xin-Qiang Li,\note{Corresponding author.}}
\author[c]{Yan Shi}
\author[b,c]{and Ya-Dong Yang}

\affiliation[a]{School of Computational Science and Electronics, Hunan Institute of Engineering, Xiangtan 411104, China}
\affiliation[b]{Institute of Particle and Nuclear Physics, Henan Normal University, Xinxiang 453007, China}
\affiliation[c]{Institute of Particle Physics and Key Laboratory of Quark and Lepton Physics~(MOE), Central China Normal University, Wuhan, Hubei 430079, China}
\affiliation[d]{School of Physics and Electronics, Hunan University, Changsha 410082, China}
\affiliation[e]{Center for High Energy Physics, Peking University, Beijing 100871, China}

\emailAdd{dengweijun@hnie.edu.cn}
\emailAdd{caifangmin@htu.edu.cn}
\emailAdd{xqli@mail.ccnu.edu.cn}
\emailAdd{shiyan@mails.ccnu.edu.cn}
\emailAdd{yangyd@ccnu.edu.cn}

\abstract{Motivated by the deviations observed between the experimental measurements and the Standard Model (SM) predictions of the branching ratios of $\bar{B}_s^0\to D_s^+ \pi^-$ and $\bar{B}_d^0\to D^+ K^-$ decays, we revisit the two-body hadronic $B_{c}^{-}\to J/\psi(\eta_{c}) L^{-}$ decays, with $L=\pi, K^{(*)}, \rho$, both within the SM and beyond. Since these processes are also mediated by the quark-level $b\to c \bar{u} d(s)$ transitions and hence dominated by the colour-allowed tree topology, the QCD factorization (QCDF) is generally expected to hold in the heavy-quark limit. Firstly, we update the SM predictions of the observables of these decays by including the nonfactorizable vertex corrections to the hadronic matrix elements of the SM four-quark operators up to the next-to-next-to-leading order (NNLO) in $\alpha_s$. It is found that, relative to the leading-order results, the branching ratios of these decays up to the next-to-leading-order (NLO) and NNLO corrections are always enhanced, with a relative amount given by $\delta_{\text{NLO}} = (\mathcal{B}^\text{NLO}-\mathcal{B}^\text{LO})/\mathcal{B}^\text{LO} \approx +6\%$ and $\delta_{\text{NNLO}} = (\mathcal{B}^\text{NNLO}-\mathcal{B}^\text{LO})/\mathcal{B}^\text{LO} \approx +9\%$, respectively. To minimize the uncertainties brought by the CKM matrix element $V_{cb}$ as well as the $B_{c}\to J/\psi(\eta_c)$ and $B_{(s)}\to D_{(s)}^{(*)}$ transition form factors, we construct ratios of the nonleptonic decay rates with respect to the corresponding differential semileptonic decay rates evaluated at $q^2=m_L^2$ (for $R_{J/\psi(\eta_{c}) L}$ and $R_{(s)L}^{(\ast)}$) or integrated over the whole $q^2$ range (for $R_{\pi/\mu\nu_{\mu}}$), which are then used to constrain the model-independent new physics (NP) Wilson coefficients. After considering the latest Belle data on $\bar{B}^0\to D^{(*)+} \pi(K)^{-}$ decays and the updated fitting results of the $B_{(s)}\to D_{(s)}^{(*)}$ transition form factors, we find that the deviations can still be explained by the NP four-quark operators with $(1+\gamma_{5}) \otimes (1-\gamma_{5})$ and $(1+\gamma_{5}) \otimes (1+\gamma_{5})$ structures, while the solution with $\gamma^\mu (1+\gamma_{5}) \otimes \gamma_\mu (1-\gamma_{5})$ structure does not work anymore, under the combined constraints from the ratios $R_{(s)L}^{(\ast)}$ at the $2\sigma$ level. Furthermore, the ratio $R_{\pi/\mu\nu_{\mu}}$, once measured precisely, could provide complementary constraint on these NP Wilson coefficients allowed by $R_{(s)L}^{(\ast)}$. With the large $B_c$ events expected at the LHC, we hope to obtain more precise measurements of these observables, which can be exploited to discriminate the different NP scenarios responsible for the deviations observed in the two-body nonleptonic $B$-meson decays.}

\begin{document}
\maketitle
\flushbottom

\section{Introduction}
\label{sec:intro}

The $B_{c}^-$ meson is the ground-state composed of two relatively heavy quarks with different flavours (i.e., a bottom quark $b$ and a charm anti-quark $\bar{c}$), and it was first discovered through the semileptonic $B_c^- \to J/\psi \ell^- \bar{\nu}_\ell$ decay in 1998 by the CDF collaboration~\cite{CDF:1998axz,CDF:1998ihx}. Ever since then, much effort has been made to study its properties, deepening our understanding of both QCD and weak interactions~\cite{QuarkoniumWorkingGroup:2004kpm,Brambilla:2010cs,HeavyFlavorAveragingGroupHFLAV:2024ctg,ParticleDataGroup:2024cfk}. Thanks to the high luminosity and high energy available at the Large Hadron Collider (LHC), detailed studies of the productions, decays and other properties of the $B_c^-$ meson are being highly promoted by the LHCb, ATLAS and CMS experiments~\cite{HeavyFlavorAveragingGroupHFLAV:2024ctg,ParticleDataGroup:2024cfk}. The current status and future prospects for $B_c^-$-meson studies at the LHC could be found, e.g., in refs.~\cite{Gouz:2002kk,Gao:2010zzc}. 

The $B_c^-$ meson is stable against both strong and electromagnetic interactions, because it lies below the $B\bar{D}$ threshold and contains two heavy quarks with open flavour quantum numbers. Thus, it can only decay weakly through the following three categories~\cite{Gouz:2002kk}: the $b$-quark decay with the $\bar{c}$ quark as a spectator, the $\bar{c}$-quark decay with the $b$ quark as a spectator, and the simultaneous annihilation of the $b$ and $\bar{c}$ quarks; they account for about $20\%$, $70\%$, and $10\%$ of the total decay width of the $B_c^-$ meson, respectively. All these three decay patterns are interesting, and offer new insights into the weak decay dynamics of heavy quarks~\cite{QuarkoniumWorkingGroup:2004kpm,Brambilla:2010cs}. The rich decay channels of the $B_c^-$ meson provide also an ideal platform for testing the Standard Model (SM) of particle physics and discriminating the different new physics (NP) scenarios beyond it. 

In this paper, we will focus on the exclusive two-body decays $B_{c}^-\to J/\psi(\eta_{c})L^{-}$ (with $L=\pi, K^{(*)}, \rho$), which belong to the first category and are mediated at the quark level by the tree-level $b\to c \bar{u} d(s)$ transitions. This is mainly motivated by the interesting observation that the measured branching ratios~\cite{HeavyFlavorAveragingGroupHFLAV:2024ctg,ParticleDataGroup:2024cfk} of the $\bar{B}_s^0\to D_s^+ \pi^-$ and $\bar{B}_d^0\to D^+ K^-$ decays, which are also mediated by the same quark-level $b\to c \bar{u} d(s)$ transitions, deviate significantly from the latest SM predictions~\cite{Huber:2016xod,Bordone:2020gao} based on the QCD factorization (QCDF) approach~\cite{Beneke:1999br,Beneke:2000ry,Beneke:2001ev}. Such a deviation could originate from some potentially underestimated corrections to the QCDF predictions (see e.g. refs.~\cite{Huber:2016xod,Bordone:2020gao,Beneke:2000ry,Fleischer:2010ca,Endo:2021ifc,Beneke:2021jhp,Piscopo:2023opf,Davies:2024vmv,Yaouanc:2024gjl}) or from some genuine NP effects (see e.g. refs.~\cite{Bobeth:2014rda,Brod:2014bfa,Lenz:2019lvd,Iguro:2020ndk,Cai:2021mlt,Bordone:2021cca,Fleischer:2021cct,Fleischer:2021cwb,Gershon:2021pnc,Lenz:2022pgw,Panuluh:2024poh,Atkinson:2024hqp,Meiser:2024zea}). Since all the four quark flavours in $b\to c\bar{u}d(s)$ transitions are different from each other, these tree-level decays have no pollution from penguin operators or penguin topologies. There is also no colour-suppressed tree topology in these so-called ``class-I'' decays, an old nomenclature adopted traditionally in $B$ physics~\cite{Neubert:1997uc}. For the $B_{c}^-\to J/\psi(\eta_{c})L^{-}$ as well as $\bar{B}_s^0\to D_s^+ \pi^-$ and $\bar{B}_d^0\to D^+ K^-$ decays, the power-suppressed annihilation contributions are even absent. All the above features make these nonleptonic decays theoretically clean and, as emphasized already in refs.~\cite{Huber:2016xod,Bordone:2020gao}, it is quite difficult to accommodate the deviation within the SM, even when the dominant higher-order power corrections from the higher-twist light-meson light-cone distribution amplitude (LCDA), the emission of a hard-collinear gluon from the spectator quark or from the heavy bottom and charm quarks, as well as the exchange of a soft gluon between the $\bar{B}_{(s)}\to D_{(s)}$ system and the light meson~\cite{Bordone:2020gao,Beneke:2000ry,Meiser:2024zea} are taken into account.\footnote{It must be pointed out that any reliable estimation of these power corrections is a very challenging task, and future improvements and investigations are certainly highly desirable. For example, a recent analysis carried out in the light-cone sum rule (LCSR) framework finds that the deviation could be explained within the SM, but with very large uncertainty due to the limited precision in the nonperturbative input as well as the leading-order accuracy of the analysis~\cite{Piscopo:2023opf}.} Hence, it seems well-motivated to entertain the possibility of this deviation being due to NP beyond the SM. Instead, it has been found that the observed deviation could be well explained by the model-independent NP four-quark operators with $\gam^{\mu}(1-\gam_5)\otimes\gam_{\mu} (1-\gam_5)$, $(1+\gam_5)\otimes(1-\gam_5)$ and $(1+\gam_5)\otimes(1+\gam_5)$ structures~\cite{Cai:2021mlt}. These NP effects in class-I nonleptonic $B$-meson decays would also affect collider observables, and some tension has been found between the NP explanations of the discrepancies observed in these decays and the dijet resonance searches~\cite{Bordone:2021cca}. However, specific assumptions on the flavour structures have to be made to simplify the whole analysis. It would be also instructive to further test these NP scenarios by considering other decays mediated by the same quark-level transitions. To this end, we will firstly update the analysis made in ref.~\cite{Cai:2021mlt} by considering the latest experimental data on $\bar{B}_{(s)}^0\to D_{(s)}^{(*)+} L^-$ decays~\cite{HeavyFlavorAveragingGroupHFLAV:2024ctg,ParticleDataGroup:2024cfk,LHCb:2021qbv,Belle:2021udv,Belle:2022afp} as well as the latest theoretical inputs, and then investigate the same NP effects on the $B_{c}^-\to J/\psi(\eta_{c})L^{-}$ decays. 

The two-body hadronic $B_{c}^-\to J/\psi(\eta_{c})L^{-}$ decays have been widely studied using various theoretical approaches and phenomenological models (see e.g. refs.~\cite{Issadykov:2018myx,Cheng:2021svx,Biswas:2023bqz,Dey:2025xdx,Wu:2024gcq,S:2024adt,Liu:2023kxr,Zhang:2023ypl,Nayak:2022qaq,Liu:2018kuo,Chang:2014jca,Rui:2014tpa,Xiao:2013lia,Ke:2013yka,Qiao:2012hp,Naimuddin:2012dy,Choi:2009ym,Sun:2007ei,Hernandez:2006gt,Ivanov:2006ni,Ebert:2003cn,Kiselev:2002vz,Colangelo:1999zn,AbdEl-Hady:1999jux,Anisimov:1998uk,Chang:1992pt} and references therein). Here, to analyze these decay processes, we will adopt the well-established QCDF approach~\cite{Beneke:1999br,Beneke:2000ry,Beneke:2001ev}, which provides a powerful and systematic framework for nonleptonic $B$-meson decays, allowing for the separation of perturbative and nonperturbative QCD effects in the heavy-quark limit (see ref.~\cite{Beneke:2015wfa} for a recent overview). It is noted that two of these decay modes, $B_{c}^-\to J/\psi \pi^-$ and $B_{c}^-\to \eta_c \pi^-$, have already been studied in the same framework up to the next-to-leading order (NLO) in the QCD coupling $\alpha_s$~\cite{Sun:2007ei}. We will update the SM predictions of the branching ratios of $B_{c}^-\to J/\psi(\eta_{c})L^{-}$ decays by including the next-to-next-to-leading order (NNLO) QCD correction to the hard-scattering kernels. For the hadronic matrix elements of the NP four-quark operators, on the other hand, we will include only the NLO QCD correction calculated within the QCDF framework. In order to minimize the uncertainties brought by the Cabibbo-Kobayashi-Maskawa (CKM) matrix element $V_{cb}$ as well as the $B_{c}\to J/\psi$ and $B_{c}\to \eta_c$ transition form factors, we will also consider the ratios of the nonleptonic decay rates with respect to the corresponding differential semileptonic decay rates evaluated at $q^2=m_L^2$ (for $R_{J/\psi(\eta_{c}) L}=\Gamma(B_{c}^-\to J/\psi(\eta_{c}) L^-)/d\Gamma(B_{c}^-\to J/\psi(\eta_{c}) \ell^-\bar{\nu}_{\ell})/dq^2\mid_{q^2=m_L^2}$) or integrated over the whole $q^2$ range $m_\mu^2\leq q^2 \leq (m_{B_{c}}-m_{J/\psi})^2$ (for $R_{\pi/\mu\nu_{\mu}}=\Gamma(B_{c}^{-}\to J/\psi \pi^{-})/\Gamma(B_{c}^{-}\to J/\psi \mu^{-}\bar{\nu}_{\mu})$), where $\ell$ refers to either an electron or a muon, and $q^2$ is the four-momentum squared transferred to the lepton pair~\cite{Bjorken:1988kk,Neubert:1997uc}. Interestingly, the LHCb collaboration has measured the ratio of nonleptonic and semileptonic branching fractions~\cite{LHCb:2014rck}
\begin{equation} \label{eq:LHCb-Rpimunu}
    R_{\pi/\mu\nu_{\mu}}^\mathrm{exp} = \frac{\mathcal{B}^\mathrm{exp}(B_{c}^{-}\to J/\psi \pi^{-})}{\mathcal{B}^\mathrm{exp}(B_{c}^{-}\to J/\psi \mu^{-}\bar{\nu}_{\mu})} = 0.0469 \pm 0.0028~(\text{stat}) \pm 0.0046~(\text{syst})\,,
\end{equation}
which is however at the lower end of the theoretical predictions (see e.g. refs.~\cite{Ivanov:2006ni,Ebert:2003cn,Kiselev:2002vz,Colangelo:1999zn,AbdEl-Hady:1999jux,Anisimov:1998uk,Chang:1992pt}). Under the combined constraints from $R_{\pi/\mu\nu_{\mu}}$ as well as the updated ratios $R_{(s)L}^{(*)}=\Gamma(\bar B_{(s)}^0\to D_{(s)}^{(*)+} L^-)/d\Gamma(\bar B_{(s)}^0\to D_{(s)}^{(*)+} \ell^-\bar{\nu}_{\ell})/dq^2\mid_{q^2=m_L^2}$ constructed in ref.~\cite{Cai:2021mlt}, we find that the deviations observed in the class-I nonleptonic $B$-meson decays could still be well explained by the NP four-quark operators with $(1+\gamma_{5})\otimes(1-\gamma_{5})$ and $(1+\gamma_{5})\otimes(1+\gamma_{5})$ structures at the $2\sigma$ level. It is worth noting that the ratio $R_{\pi/\mu\nu_{\mu}}$, once measured precisely, can further narrow down the NP parameter space allowed by the ratios $R_{(s)L}^{(*)}$, and hence help us to probe the various NP scenarios responsible for these class-I $B$-meson decays.

Our paper is organized as follows. In section~\ref{sec:theory}, after introducing the most general effective weak Hamiltonian describing the quark-level $b\to c\bar{u}d(s)$ transitions, we describe the calculations of the hadronic matrix elements of the SM and NP four-quark operators within the QCDF framework, and define the branching fractions and the relevant ratios relevant to $B_{c}^-\to J/\psi(\eta_{c})L^{-}$ decays. In section~\ref{sec:Numerical analysis}, we firstly present the updated SM predictions of these observables, and then discuss the NP effects on these decays in a model-independent setup. Our conclusions are finally made in section~\ref{sec:conclusions}. %For convenience, the updated allowed ranges of the NP Wilson coefficients $C_i(m_b)$ under the individual and combined constraints from the ratios $R_{(s)L}^{(\ast)}$ are presented in the appendix.

\section{Theoretical framework}
\label{sec:theory}

As demonstrated in ref.~\cite{Sun:2007ei}, the QCD factorization for the two-body hadronic $B_{c}^-\to J/\psi(\eta_{c})L^{-}$ decays is generally expected to hold in the heavy-quark limit, and the hadronic matrix elements of the SM four-quark operators have been calculated up to the NLO in $\alpha_s$ by including the nonfactorizable one-loop vertex corrections within the QCDF approach~\cite{Beneke:1999br,Beneke:2000ry,Beneke:2001ev}. Furthermore, a factorization formula for the decay amplitudes of these processes valid up to the one-loop level~\cite{Qiao:2012hp} has been established in the nonrelativistic QCD (NRQCD) approach~\cite{Bodwin:1994jh,Brambilla:2004jw}. Here we will follow these ideas to analyze the $B_{c}^-\to J/\psi(\eta_{c})L^{-}$ decays by including the nonfactorizable vertex corrections to the hadronic matrix elements of the SM and NP four-quark operators up to the NNLO and NLO in $\alpha_s$ respectively, within the QCDF framework.   

\subsection{Effective weak Hamiltonian}
\label{subsec:effectiveD}

Just like the class-I $\bar B_{(s)}^0\to D_{(s)}^{(*)+} L^-$ decays, the $B_{c}^-\to J/\psi(\eta_{c})L^{-}$ decays are also mediated by the quark-level $b\to c\bar{u}d(s)$ transitions. Once the top quark, the gauge bosons $W^\pm$ and $Z^0$, the Higgs boson, as well as all other heavy degrees of freedom present in any extension of the SM are integrated out, the decay amplitudes of these processes can be computed most conveniently in the framework of Weak Effective Theory (WET)~\cite{Buchalla:1995vs,Jenkins:2017jig,Aebischer:2017gaw}. Specifically, the most general effective weak Hamiltonian for the quark-level $b\to c\bar{u}d(s)$ transitions can be written as\footnote{Here we assume that the NP scale $\mu_0$ satisfies the condition $\mu_0\gg m_b$, ensuring therefore that all the NP effects can be accounted for by such a low-energy effective weak Hamiltonian.}
\begin{align} \label{eq:Hamiltonian}
  \mH_\text{eff} &= \frac{G_F}{\sqrt{2}}\,V_{cb}V^*_{uq}\,\bigg\{\sum_{i}\mC_i(\mu)\mQ_i(\mu) +\sum_{i,j}\Big[C_{i}^{VLL}(\mu)\mQ_{i}^{VLL}(\mu) + C_{i}^{VLR}(\mu)\mQ_{i}^{VLR}(\mu) \nn \\[0.1cm] 
  & \hspace{2.6cm} + C_{j}^{SLL}(\mu)\mQ_{j}^{SLL}(\mu) + C_{i}^{SLR}(\mu) \mQ_{i}^{SLR}(\mu) + (L\leftrightarrow R)\Big]\bigg\} + \text{h.c.} \,, 
\end{align}
where $G_F$ is the Fermi constant, and $V_{cb}V^*_{uq}$ the product of the CKM matrix elements, with $q=d$ or $s$. It should be noted that, besides being the low-energy effective theory of the SM weak interactions, the WET can also be regarded as the low-energy effective theory of any quantum field theory that shares the same field content and gauge symmetry with the SM below the electroweak scale. Thus, the WET has also been called the ``Low-Energy Effective Field Theory''~\cite{Jenkins:2017jig}. Such a framework facilitates scale factorization and allows to resum large logarithms of the form $\ln\mu_0/m_b$ (for a recent review, see e.g. ref.~\cite{Buras:2011we}). 

In eq.~\eqref{eq:Hamiltonian}, $\mQ_{i}$ ($i=1,2$) represent the two four-quark current-current operators within the SM that are given, respectively, as~\cite{Buchalla:1995vs}
\begin{equation} \label{eq:SM-operator}
\mQ_{1} = \bigl[\overline{c}_{\alpha}\gamma^{\mu}(1-\gam_5)b_{\beta}\bigr] \bigl[\overline{q}_{\beta}\gamma_{\mu}(1-\gam_5)u_{\alpha}\bigr]\,, \quad
\mQ_{2} = \bigl[\overline{c}_{\alpha}\gamma^{\mu}(1-\gam_5)b_{\alpha}\bigr] \bigl[\overline{q}_{\beta}\gamma_{\mu}(1-\gam_5)u_{\beta}\bigr]\,,
\end{equation}
while the remaining ones denote the full set of twelve linearly-independent four-quark operators that can contribute, either directly or through operator mixing, to the quark-level $b\to c \bar u d(s)$ transitions. The latter can be further partitioned into eight independent sectors that do not mix with each other under renormalization up to order of $G_F\sim g^2/m_W^2$. Explicitly, the first four sectors are given, respectively, as~\cite{Buras:2000if,Buras:2012gm}\footnote{Here the NP four-quark operators are given in a Fierz transformed version of the Buras-Misiak-Urban basis~\cite{Buras:2000if}. They can also be expressed in the so-called Bern basis~\cite{Aebischer:2017gaw}, and the exchange between these two bases could be found, e.g., in ref.~\cite{Meiser:2024zea}. }
\begin{align} \label{eq:NP-operator}
\mQ_{1}^{VLL}&=\bigl[\bar{c}_{\alpha}\gamma^{\mu}(1-\gam_5)b_{\beta}\bigr] 
\bigl[\bar{q}_{\beta}\gamma_{\mu}(1-\gam_5)u_{\alpha}\bigr], \hspace{-0.2cm} &
\mQ_{2}^{VLL}&=\bigl[\bar{c}_{\alpha}\gamma^{\mu}(1-\gam_5)b_{\alpha}\bigr] 
\bigl[\bar{q}_{\beta}\gamma_{\mu}(1-\gam_5)u_{\beta}\bigr], \nn \\[0.3cm]
\mQ_{1}^{VLR}&=\bigl[\bar{c}_{\alpha}\gamma^{\mu}(1-\gam_5)b_{\beta}\bigr] 
\bigl[\bar{q}_{\beta}\gamma_{\mu}(1+\gam_5)u_{\alpha}\bigr], \hspace{-0.2cm} &
\mQ_{2}^{VLR}&=\bigl[\bar{c}_{\alpha}\gamma^{\mu}(1-\gam_5)b_{\alpha}\bigr]
\bigl[\bar{q}_{\beta}\gamma_{\mu}(1+\gam_5)u_{\beta}\bigr], \nn \\[0.3cm]
\mQ_{1}^{SLL}&=\bigl[\bar{c}_{\alpha}(1-\gam_5)b_{\beta}\bigr] 
\bigl[\bar{q}_{\beta}(1-\gam_5)u_{\alpha}\bigr], \hspace{-0.2cm} & 
\mQ_{3}^{SLL}&=\bigl[\bar{c}_{\alpha}\sigma^{\mu\nu}(1-\gam_5)b_{\beta}\bigr] 
\bigl[\bar{q}_{\beta}\sigma_{\mu\nu}(1-\gam_5)u_{\alpha}\bigr], \nn \\[0.15cm]
\mQ_{2}^{SLL}&=\bigl[\bar{c}_{\alpha}(1-\gam_5)b_{\alpha}\bigr]
\bigl[\bar{q}_{\beta}(1-\gam_5)u_{\beta}\bigr], \hspace{-0.2cm} & 
\mQ_{4}^{SLL}&=\bigl[\bar{c}_{\alpha}\sigma^{\mu\nu}(1-\gam_5)b_{\alpha}\bigr]
\bigl[\bar{q}_{\beta}\sigma_{\mu\nu}(1-\gam_5)u_{\beta}\bigr], \nn \\[0.3cm]
\mQ_{1}^{SLR}&=\bigl[\bar{c}_{\alpha}(1-\gam_5)b_{\beta}\bigr]
\bigl[\bar{q}_{\beta}(1+\gam_5)u_{\alpha}\bigr], \hspace{-0.2cm} &
\mQ_{2}^{SLR}&=\bigl[\bar{c}_{\alpha}(1-\gam_5)b_{\alpha}\bigr]
\bigl[\bar{q}_{\beta}(1+\gam_5)u_{\beta}\bigr],
\end{align}
where $\alpha$ and $\beta$ are the colour indices, and $\sigma^{\mu\nu}=\frac{1}{2}[\gamma^\mu,\gamma^\nu]$. The NP operators belonging to the remaining four chirality-flipped counterparts, $VRR$, $VRL$, $SRR$ and $SRL$, are obtained from eq.~\eqref{eq:NP-operator} by making the interchanges of $(1\mp\gam_5)\leftrightarrow (1\pm\gam_5)$. The short-distance Wilson coefficients $\mC_{i}(\mu)$ and $C_{i}(\mu)$ in eq.~\eqref{eq:Hamiltonian} can be calculated using the renormalization group (RG) improved perturbation theory~\cite{Buchalla:1995vs,Buras:2011we}. Explicit expressions of the SM parts, $\mC_{i}(\mu)$, up to next-to-next-to-leading logarithmic accuracy in QCD can be found, e.g., in ref.~\cite{Gorbahn:2004my} and will be used throughout this paper. For the NP parts $C_{i}(\mu)$, on the other hand, one can easily obtain the next-to-leading logarithmic results of $C_{i}(\mu_b)$ evaluated at the characteristic scale $\mu_b\simeq m_b$ appropriate for the nonleptonic $B$-meson decays, by solving the RG equations satisfied by them, based on the one- and two-loop QCD anomalous dimension matrices (ADMs) of the full set of NP four-quark operators~\cite{Buras:2000if}, as well as the $\mathcal{O}(\alpha_s)$ corrections to the matching conditions for $C_{i}(\mu_0)$ evaluated at the NP scale $\mu_0$~\cite{Buras:2012gm}. 
  
\subsection{Calculation of the hadronic matrix elements}
\label{subsec:HSK}

To obtain the QCD amplitudes of $B_{c}^-\to J/\psi(\eta_{c})L^{-}$ decays, our next task is to calculate the hadronic matrix elements of the local four-quark operators present in eq.~\eqref{eq:Hamiltonian}. To this end, we will adopt the QCDF approach~\cite{Beneke:1999br,Beneke:2000ry,Beneke:2001ev}, according to which the following factorization formula is established in the heavy-quark limit:
\begin{equation}\label{eq:QCDF}
\langle J/\psi(\eta_{c}) L^-|\mQ_{i}(\mu)|B_{c}^{-}\rangle = \sum_{j} F_{j}^{B_{c}\to J/\psi(\eta_{c})}(m_{L}^{2})\,\int_{0}^{1} du\,T_{ij}(u,\mu)\,\Phi_{L}(u,\mu) + \mathcal{O}(\Lambda_\mathrm{QCD}/m_b)\,,
\end{equation}
where $F_{j}^{B_{c}\to J/\psi(\eta_{c})}$ is the $B_{c}\to J/\psi(\eta_{c})$ transition form factor in full QCD, and $\Phi_{L}(u,\mu)$ the LCDA of the light meson $L^-$. They encode all the long-distance strong interaction effects involved in the two-body nonleptonic decays, and can be extracted from experimental data or calculated by using the nonperturbative methods like QCD sum rules~\cite{Colangelo:1992cx,Kiselev:1999sc,Leljak:2019eyw,Wu:2024gcq,Bordone:2022drp,Straub:2015ica,Dimou:2012un,Ball:2006nr,Ball:2005vx,Ball:1996tb} and/or lattice QCD~\cite{Harrison:2025kxm,Harrison:2020gvo,Harrison:2020nrv,Colquhoun:2016osw,Arthur:2010xf,Bali:2019dqc,Braun:2016wnx,LatticeParton:2022zqc}. The $B_{c}\to J/\psi(\eta_{c})$ transition form factors can also be calculated within the NRQCD framework by including the higher-order perturbative and relativistic corrections~\cite{Bell:2006tz,Bell:2005gw,Qiao:2011yz,Qiao:2012vt,Zhu:2017lqu,Shen:2021dat,Tao:2022yur,Colangelo:2022lpy}.

The QCDF formula of eq.~\eqref{eq:QCDF} is based on the observation that, in the heavy-quark limit of $m_{b,c}\to \infty$ for fixed $m_c/m_b$, the light quark-antiquark pair $\bar{u}d(s)$ created from the weak $b\to c \bar u d(s)$ transitions must be highly energetic and collinear to form the light meson $L^-$, which is predominantly produced as a compact object with small transverse extension~\cite{Beneke:1999br,Beneke:2000ry,Beneke:2001ev}. Thus, for the strong interactions between the emitted light meson $L^-$ and the $B_{c} \to J/\psi(\eta_{c})$ system, only hard gluons survive at the leading power in $\Lambda_\mathrm{QCD}/m_b$, while soft and collinear gluon effects are power-suppressed, which is a technical manifestation of Bjorken's colour-transparency argument~\cite{Bjorken:1988kk}. Furthermore, as demonstrated in ref.~\cite{Qiao:2012hp}, the factorizable hard-spectator scattering contributions account for more than $85\%$ of the total results at the NLO accuracy in the heavy-quark limit. This implies that the nonfactorizable hard-spectator scattering contributions play only a minor role in these class-I $B_{c}^-\to J/\psi(\eta_{c})L^{-}$ decays, as is generally expected in two-body hadronic $B$-meson decays dominated by the colour-allowed tree topology~\cite{Beneke:1999br,Beneke:2000ry,Beneke:2001ev,Beneke:2003zv,Beneke:2005vv,Pilipp:2007mg,Kivel:2006xc,Beneke:2009ek,Bell:2007tv,Bell:2009nk,Huber:2016xod}. We are, therefore, motivated to consider only the nonfactorizable vertex corrections, which are totally dominated by hard gluon interactions and can be calculated perturbatively order by order in $\alpha_s$. Explicitly, we can write the hard-scattering kernels $T_{ij}(u,\mu)$ in eq.~\eqref{eq:QCDF} as
\begin{equation}
    T_{ij}(u,\mu) = T_{ij}^{(0)}(u,\mu) +\frac{\alpha_s}{4\pi}\, T_{ij}^{(1)}(u,\mu) +\left(\frac{\alpha_s}{4\pi}\right)^2\, T_{ij}^{(2)}(u,\mu) + \cdots\,,
\end{equation}
where the coefficients $T_{ij}^{(0,1,2)}$ receive only contributions from scales of $\mathcal{O}(m_b)$ in the heavy-quark limit, and can be obtained from the leading-order (LO), NLO and NNLO quark-level Feynman diagrams shown in figures~\ref{fig:LO}--\ref{fig:vertex_NNLO}, respectively. 

%%%%%%%%%%%%%%%%%%%%%%%%%%%%%%%%%%%%%%%%%%%%%%%%%%%%%%
\begin{figure}[t]\fontsize{7.0}{10}
	\begin{center}
		 \begin{tikzpicture}[line width=1.0pt, scale=1.52, >=Stealth]
		 	\begin{scope}
		 		\draw[fermionbar](0:1)--(0,0);
		 		\node at (0.5,-0.12) {$b$};
		 		\node at (1.0,0.0) {$\blacksquare$};
		 	\end{scope}ij
		 	\begin{scope}[shift={(1,0)}]
		 		\draw[fermionbar](0:1)--(0,0);
		 		\node at (0.5,-0.12) {$c$};
		 		\draw[fermion](60:1)--(0,0);
		 		\draw[fermionbar](120:1)--(0,0);
		 		\node at (-0.5,0.6) {$q$};
		 		\node at (0.5,0.6) {$\bar{u}$};
		 	\end{scope}ij
		 	\begin{scope}[shift={(0.0,-0.7)}]
		 		\draw[fermion](0:2)--(0,0);
		 		\node at (1,-0.12) {$\bar{c}$};
		 	\end{scope}ij
		 \end{tikzpicture}
        %\centering
        %\includegraphics[width=0.25\textwidth]{figure/LOdiag.pdf}
		\caption{LO quark-level Feynman diagram contributing to the hard-scattering kernels $T_{ij}(u,\mu)$, where the local four-quark operator is represented by the black square. \label{fig:LO}}
	\end{center}
\end{figure}
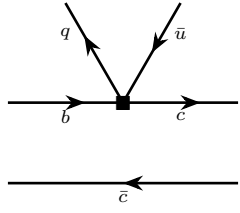
%%%%%%%%%%%%%%%%%%%%%%%%%%%%%%%%%%%%%%%%%%%%%%%%%%%%%%

%%%%%%%%%%%%%%%%%%%%%%%%%%%%%%%%%%%%%%%%%%%%%%%%%%%%%%
\begin{figure}[t]\fontsize{7.0}{10}
	 \begin{center}
	 	\begin{tikzpicture}[line width=1.0pt, scale=1.52, >=Stealth]
	 		\begin{scope}
	 			\draw[fermionbar](0:1)--(0,0);
	 			\draw[gluon](0.3,0)--(0.78,0.40);
	 			\node at (0.5,-0.12) {$b$};
	 			\node at (1.0,0.0) {$\blacksquare$};
	 		\end{scope}ij
	 		\begin{scope}[shift={(1,0)}]
	 			\draw[fermionbar](0:1)--(0,0);
	 			\node at (0.5,-0.12) {$c$};
	 			\draw[fermion](60:1)--(0,0);
	 			\draw[fermionbar](120:1)--(0,0);
	 			\node at (-0.5,0.6) {$q$};
	 			\node at (0.5,0.6) {$\bar{u}$};
	 		\end{scope}ij
	 		\begin{scope}[shift={(0.0,-0.7)}]
	 			\draw[fermion](0:2)--(0,0);
	 			\node at (1,-0.12) {$\bar{c}$};
	 		\end{scope}ij
			
	 		\begin{scope}[shift={(2.5,0)}]
	 			\draw[fermionbar](0:1)--(0,0);
	 			\draw[gluon](0.3,0)--(1.235,0.40);
	 			\node at (0.5,-0.12) {$b$};
	 			\node at (1.0,0.0) {$\blacksquare$};
	 		\end{scope}ij
	 		\begin{scope}[shift={(3.5,0)}]
	 			\draw[fermionbar](0:1)--(0,0);
	 			\node at (0.5,-0.12) {$c$};
	 			\draw[fermion](60:1)--(0,0);
	 			\draw[fermionbar](120:1)--(0,0);
	 			\node at (-0.5,0.6) {$q$};
	 			\node at (0.5,0.6) {$\bar{u}$};
	 		\end{scope}ij
	 		\begin{scope}[shift={(2.5,-0.7)}]
	 			\draw[fermion](0:2)--(0,0);
	 			\node at (1,-0.12) {$\bar{c}$};
	 		\end{scope}ij
			
	 		\begin{scope}[shift={(5,0)}]
	 			\draw[fermionbar](0:1)--(0,0);
	 			\node at (0.5,-0.12) {$b$};
	 			\node at (1.0,0.0) {$\blacksquare$};
	 		\end{scope}ij
	 		\begin{scope}[shift={(6.72,0)}]
	 			\draw[gluon](156:1.05)--(0.0,0);
	 		\end{scope}ij
	 		\begin{scope}[shift={(6,0)}]
	 			\draw[fermionbar](0:1)--(0,0);
	 			\node at (0.5,-0.12) {$c$};
	 			\draw[fermion](60:1)--(0,0);
	 			\draw[fermionbar](120:1)--(0,0);
	 			\node at (-0.5,0.6) {$q$};
	 			\node at (0.5,0.6) {$\bar{u}$};
	 		\end{scope}ij
	 		\begin{scope}[shift={(5,-0.7)}]
	 			\draw[fermion](0:2)--(0,0);
	 			\node at (1,-0.12) {$\bar{c}$};
	 		\end{scope}ij
			
	 		\begin{scope}[shift={(7.5,0)}]
	 			\draw[fermionbar](0:1)--(0,0);
	 			\node at (0.5,-0.12) {$b$};
	 			\node at (1.0,0.0) {$\blacksquare$};
	 		\end{scope}ij
	 		\begin{scope}[shift={(9.22,0)}]
	 			\draw[gluon](137:0.64)--(0,0);
	 		\end{scope}ij
	 		\begin{scope}[shift={(8.5,0)}]
	 			\draw[fermionbar](0:1)--(0,0);
	 			\node at (0.5,-0.12) {$c$};
	 			\draw[fermion](60:1)--(0,0);
	 			\draw[fermionbar](120:1)--(0,0);
	 			\node at (-0.5,0.6) {$q$};
	 			\node at (0.5,0.6) {$\bar{u}$};
	 		\end{scope}ij
	 		\begin{scope}[shift={(7.5,-0.7)}]
	 			\draw[fermion](0:2)--(0,0);
	 			\node at (1,-0.12) {$\bar{c}$};
	 		\end{scope}ij
	 	\end{tikzpicture}
        %\centering
        %\includegraphics[width=0.98\textwidth]{figure/NLOdiag.pdf}
		\caption{Nonfactorizable vertex corrections to the hard-scattering kernels $T_{ij}(u,\mu)$ at the NLO in $\alpha_s$, where the other captions are the same as in figure~\ref{fig:LO}. \label{fig:vertex_NLO}}
	 \end{center}
\end{figure}
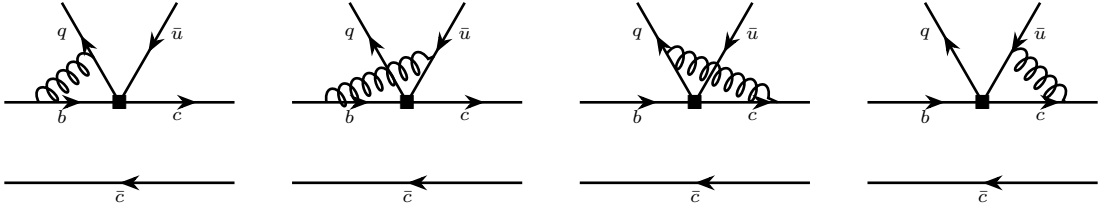
%%%%%%%%%%%%%%%%%%%%%%%%%%%%%%%%%%%%%%%%%%%%%%%%%%%%%

%%%%%%%%%%%%%%%%%%%%%%%%%%%%%%%%%%%%%%%%%%%%%%%%%%%%%%
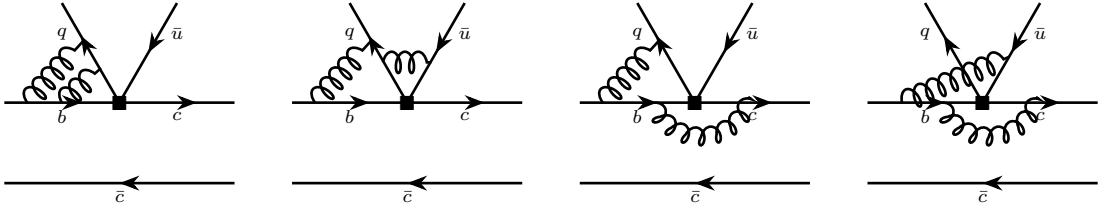
\begin{figure}[t]\fontsize{7.0}{10}
	 \begin{center}
	 	\begin{tikzpicture}[line width=1.0pt, scale=1.52, >=Stealth]
	 		\begin{scope}
	 			\draw[fermionbar](0:1)--(0,0);
	 			\draw[gluon](0.2,0)--(0.68,0.53);
	 			\draw[gluon](0.5,0)--(0.84,0.30);
	 			\node at (0.5,-0.12) {$b$};
	 			\node at (1.0,0.0) {$\blacksquare$};
	 		\end{scope}ij
	 		\begin{scope}[shift={(1,0)}]
	 			\draw[fermionbar](0:1)--(0,0);
	 			\node at (0.5,-0.12) {$c$};
	 			\draw[fermion](60:1)--(0,0);
	 			\draw[fermionbar](120:1)--(0,0);
	 			\node at (-0.5,0.6) {$q$};
	 			\node at (0.5,0.6) {$\bar{u}$};
	 		\end{scope}ij
	 		\begin{scope}[shift={(0.0,-0.7)}]
	 			\draw[fermion](0:2)--(0,0);
	 			\node at (1,-0.12) {$\bar{c}$};
	 		\end{scope}ij
			
	 		\begin{scope}[shift={(2.5,0)}]
	 			\draw[fermionbar](0:1)--(0,0);
	 			\draw[gluon](0.2,0)--(0.68,0.53);
	 			\node at (0.5,-0.12) {$b$};
	 			\node at (1.0,0.0) {$\blacksquare$};
	 		\end{scope}ij
	 		\begin{scope}[shift={(3.5,0)}]
	 		    \draw[gluon](-0.210,0.35)--(0.210,0.35);
	 			\draw[fermionbar](0:1)--(0,0);
	 			\node at (0.5,-0.12) {$c$};
	 			\draw[fermion](60:1)--(0,0);
	 			\draw[fermionbar](120:1)--(0,0);
	 			\node at (-0.5,0.6) {$q$};
	 			\node at (0.5,0.6) {$\bar{u}$};
	 		\end{scope}ij
	 		\begin{scope}[shift={(2.5,-0.7)}]
	 			\draw[fermion](0:2)--(0,0);
	 			\node at (1,-0.12) {$\bar{c}$};
	 		\end{scope}ij
			
	 		\begin{scope}[shift={(5,0)}]
	 		    \draw[gluon](0.2,0)--(0.68,0.53);
	 			\draw[fermionbar](0:1)--(0,0);
	 			\node at (0.5,-0.12) {$b$};
	 			\node at (1.0,0.0) {$\blacksquare$};
	 		\end{scope}ij
	 		\begin{scope}[shift={(6,0)}]
	 			\draw[fermionbar](0:1)--(0,0);
	 			\node at (0.5,-0.12) {$c$};
	 			\draw[fermion](60:1)--(0,0);
	 			\draw[fermionbar](120:1)--(0,0);
	 			\node at (-0.5,0.6) {$q$};
	 			\node at (0.5,0.6) {$\bar{u}$};
	 			\draw[style={decorate,draw=black,decoration={coil,aspect=0.7,amplitude=3.0pt, segment length=5.7pt}}] (-0.35,0) arc [start angle=200, end angle=347, radius=0.425];
	 		\end{scope}ij
	 		\begin{scope}[shift={(5,-0.7)}]
	 			\draw[fermion](0:2)--(0,0);
	 			\node at (1,-0.12) {$\bar{c}$};
	 		\end{scope}ij
			
	 		\begin{scope}[shift={(7.5,0)}]
	 		    \draw[gluon](0.3,0)--(1.235,0.40);
	 			\draw[fermionbar](0:1)--(0,0);
	 			\node at (0.5,-0.12) {$b$};
	 			\node at (1.0,0.0) {$\blacksquare$};
	 		\end{scope}ij
	 		\begin{scope}[shift={(8.5,0)}]
	 			\draw[fermionbar](0:1)--(0,0);
	 			\node at (0.5,-0.12) {$c$};
	 			\draw[fermion](60:1)--(0,0);
	 			\draw[fermionbar](120:1)--(0,0);
	 			\node at (-0.5,0.6) {$q$};
	 			\node at (0.5,0.6) {$\bar{u}$};
	 			\draw[style={decorate,draw=black,decoration={coil,aspect=0.7,amplitude=3.0pt, segment length=5.7pt}}] (-0.35,0) arc [start angle=200, end angle=347, radius=0.425];
	 		\end{scope}ij
	 		\begin{scope}[shift={(7.5,-0.7)}]
	 			\draw[fermion](0:2)--(0,0);
	 			\node at (1,-0.12) {$\bar{c}$};
	 		\end{scope}ij
	 	\end{tikzpicture}
    %    \centering
    %    \includegraphics[width=0.98\textwidth]{figure/NNLOdiag.pdf}
		\caption{Sample of nonfactorizable vertex diagrams that contribute to the hard-scattering kernels $T_{ij}(u,\mu)$ at the NNLO in $\alpha_s$, where the other captions are the same as in figure~\ref{fig:LO}. \label{fig:vertex_NNLO}}
	\end{center}
\end{figure}
%%%%%%%%%%%%%%%%%%%%%%%%%%%%%%%%%%%%%%%%%%%%%%%%%%%%%%

The procedure to calculate the quark-level Feynman diagrams shown in figures~\ref{fig:LO}--\ref{fig:vertex_NNLO} in the QCDF approach has been detailed in refs.~\cite{Beneke:2000ry,Huber:2016xod} for the SM current-current operators up to the NNLO in $\alpha_s$, and in refs.~\cite{Cai:2021mlt,Meiser:2024zea} for the full set of twelve linearly-independent NP four-quark operators up to the NLO in $\alpha_s$. Since these class-I $\bar B_{(s)}^0\to D_{(s)}^{(*)+} L^-$ and $B_{c}^-\to J/\psi(\eta_{c})L^{-}$ decays are all mediated by the same quark-level $b\to c\bar u d(s)$ transitions, it is easy to check that they share the same hard-scattering kernels $T_{ij}(u,\mu)$ as given in refs.~\cite{Beneke:2000ry,Huber:2016xod,Cai:2021mlt,Meiser:2024zea}.\footnote{The recent study~\cite{Meiser:2024zea} includes also the three-particle contributions to the hard-scattering kernels at the LO in $\alpha_s$, for both the SM and the complete set of WET operators, where an additional collinear gluon will hadronize with the $\bar u d(s)$ pair into the final-state light meson $L^-$.} Thus, we will repeat neither these calculations nor the analytical results of $T_{ij}(u,\mu)$. After performing the convolution over the light-cone momentum fraction $u$ of the valence quark inside the light meson $L^-$ in eq.~\eqref{eq:QCDF}, we have a formally perturbative expansion of the colour-allowed tree amplitudes $a_1(J/\psi(\eta_{c})L^{-})$~\cite{Cai:2021mlt,Meiser:2024zea}:
\begin{equation} \label{eq:a1_effective}
a_{1}(J/\psi(\eta_{c})L^{-}) = a_{1}^{(0)}(J/\psi(\eta_{c})L^{-}) + \frac{\alpha_s}{4\pi}\,a_{1}^{(1)}(J/\psi(\eta_{c})L^{-}) + \left(\frac{\alpha_s}{4\pi}\right)^2\,a_{1}^{(2)}(J/\psi(\eta_{c})L^{-}) + \cdots\,,
\end{equation} 
where $a_{1}^{(0)}(J/\psi(\eta_{c})L^{-})$ is just given in terms of a linear combination of the short-distance Wilson coefficients $\mC_{i}(\mu)$ and $C_{i}(\mu)$, because the LO hard-scattering kernels $T_{ij}^{(0)}(u,\mu)$ are independent of the momentum fraction $u$ and the convolution integrals in eq.~\eqref{eq:QCDF} reduce to the normalization conditions of the light-meson LCDA $\Phi_{L}(u,\mu)$. In this case, the QCDF formula reproduces the well-known naive factorization result~\cite{Bauer:1986bm,Fakirov:1977ta}. Starting at the NLO in $\alpha_s$, on the other hand, the hard-scattering kernels $T_{ij}^{(n)}(u,\mu)$ for $n\geq 1$ will display a nontrivial dependence on the momentum fraction $u$, making the effective coefficients $a_{1}^{(n)}(J/\psi(\eta_{c})L^{-})$ depend also on the specific meson $L^-$ through its LCDA $\Phi_{L}(u,\mu)$. Explicit expressions of these convoluted kernels with $\Phi_{L}(u,\mu)$ expanded in a basis of Gegenbauer polynomials $C_k^{3/2}(x)$ up to the second Gegenbauer moment can be found in refs.~\cite{Huber:2016xod,Cai:2021mlt}. %Especially, for convenience, we have attached the full SM result up to the NNLO in $\alpha_s$ in electronic form to the arXiv submission of ref.~\cite{Huber:2016xod}.

Finally, it should be emphasized that our calculations of the hadronic matrix elements of these four-quark operators are performed in the naive dimensional regularization scheme with anti-commuting $\gamma_5$ in $D=4-2\epsilon$ dimensions, which matches exactly the one used for evaluations of the short-distance Wilson coefficients $\mC_{i}(\mu)$ and $C_{i}(\mu)$~\cite{Buras:2012gm,Buras:2000if,Gorbahn:2004my}. This is necessary to ensure the renormalization scheme and scale independence of the NP (SM) decay amplitudes up to the (N)NLO in $\alpha_s$.

\subsection{Branching fractions and the ratios \texorpdfstring{$R_{J/\psi(\eta_{c}) L}$}{R[J/psi(eta c)L]} and \texorpdfstring{$R_{\pi/\mu\nu_{\mu}}$}{R[pi/mu nu mu]}}
\label{subsec:OBS}

In terms of the colour-allowed tree coefficients $a_1(J/\psi(\eta_{c})L^{-})$ defined in eq.~\eqref{eq:a1_effective}, the decay amplitudes of $B_{c}^-\to J/\psi(\eta_{c})L^{-}$ decays can then be written as
\begin{align}
 \mathcal{A}(B_{c}^-\to J/\psi(\eta_{c})L^{-}) & = \langle J/\psi(\eta_{c}) L^-|\mH_\text{eff}|B_{c}^{-}\rangle \nn\\[0.15cm]
  & \hspace{-3.2cm} = \frac{G_F}{\sqrt{2}}\,V_{cb}V^*_{uq}\,a_{1}(J/\psi(\eta_{c})L^{-})\,\langle L^-|\bar{d}(\bar{s})\gamma^{\mu}(1-\gamma_{5})u|0\rangle\,\langle J/\psi(\eta_{c}) |\bar{c}\gamma_{\mu}(1-\gamma_{5})b|B_{c}^{-}\rangle\,,
\end{align}
where we have used the equations of motion of quark fields to express the hadronic matrix element of any local four-quark operator in terms of that of the SM current-current operator. The latter can be further approximated as the product of the matrix elements of two bilinear quark currents in the naive factorization approach. Following the same conventions as used in ref.~\cite{Beneke:2000wa}, we can decompose and parameterize the matrix elements $\langle J/\psi|\bar{c}\gamma_{\mu}(1-\gamma_{5})b|B_{c}^{-}\rangle$ and $\langle \eta_{c}|\bar{c}\gamma_{\mu}(1- \gamma_{5})b|B_{c}^{-}\rangle$, respectively, as 
\begin{align}
\langle J/\psi(p^{\prime},\varepsilon)|\bar{c}\gamma_{\mu}b|B_{c}^{-}(p)\rangle &= \frac{2 i V(q^2)}{m_{B_{c}}+m_{J/\psi}}\epsilon_{\mu\nu\rho\sigma}\varepsilon^{*\nu}p^{\prime\rho}p^{\sigma}\,, \nn \\[0.2cm]
\langle J/\psi(p^{\prime},\varepsilon) |\bar{c}\gamma_{\mu}\gamma_{5}b|B_{c}^{-}(p)\rangle &= 2m_{J/\psi}A_{0}(q^{2})\frac{\varepsilon^{*} \cdot q}{q^{2}}q_{\mu} + \left(m_{B_{c}}+m_{J/\psi}\right)A_{1}(q^{2})\left[\varepsilon_{\mu}^{*} - \frac{\varepsilon^{*} \cdot q}{q^{2}}q_{\mu}\right] \nn \\[0.15cm] 
& - A_{2}(q^{2}) \frac{\varepsilon^{*} \cdot q}{m_{B_{c}}+m_{J/\psi}}\left[(p+p^{\prime})_{\mu}-\frac{m_{B_{c}}^2-m_{J/\psi}^2}{q^{2}}q_{\mu}\right]\,,\\[0.2cm]
\langle \eta_{c}(p^{\prime})|\bar{c}\gamma_{\mu}b|B_{c}^{-}(p)\rangle &= f_{+}(q^2)\left[(p+p^{\prime})_{\mu}-\frac{m_{B_{c}}^2-m_{\eta_{c}}^2}{q^{2}}q_{\mu}\right] + f_{0}(q^2)\frac{m_{B_{c}}^2-m_{\eta_{c}}^2}{q^{2}}q_{\mu}\,, \nn \\[0.2cm]
\langle \eta_{c}(p^{\prime})|\bar{c}\gamma_{\mu}\gamma_{5}b|B_{c}^{-}(p)\rangle &= 0\,, 
\end{align}
where $q^\mu=p^\mu-p^{\prime\mu}$, and $\epsilon_{0123}=1$. For the matrix elements $\langle L^-|\bar{d}(\bar{s})\gamma^{\mu}(1-\gamma_{5})u|0\rangle$, on the other hand, we have 
\begin{align}
\langle P^- (q)|\bar{d}(\bar{s})\gamma^{\mu}\gamma_{5}u|0\rangle &= -i f_P q^\mu\,, & 
\langle P^- (q)|\bar{d}(\bar{s})\gamma^{\mu} u|0\rangle &= 0\,, \\[0.2cm]
\langle V^- (q,\varepsilon)|\bar{d}(\bar{s})\gamma^{\mu}u|0\rangle &= -i f_V m_V \varepsilon^{*\mu}\,, &
\langle V^- (q,\varepsilon)|\bar{d}(\bar{s})\gamma^{\mu}\gamma_{5}u|0\rangle &= 0\,, 
\end{align}
where $P$ and $V$ refer to a pseudoscalar and a longitudinal vector meson respectively, and $f_L$ denotes the corresponding decay constant. Throughout this paper, we consider only the case with longitudinally polarized vector mesons in the two-body hadronic $B_{c}^-\to J/\psi V^{-}$ decays, because the two transversely polarized amplitudes are power-suppressed and cannot be calculated reliably within the QCDF framework~\cite{Beneke:2006hg,Cheng:2008gxa,Bartsch:2008ps,Beneke:2009eb}. As expected by angular momentum conservation, the final-state vector mesons must be also longitudinally polarized in the two-body hadronic $B_{c}^-\to \eta_c V^{-}$ and $B_{c}^-\to J/\psi P^{-}$ decays. 

Finally, the decay rates and branching fractions of $B_{c}^-\to J/\psi(\eta_{c})L^{-}$ decays can be written, respectively, as
\begin{align}
        \Gamma(B_{c}^{-}\to J/\psi(\eta_{c}) L^{-}) &= \frac{\vert\overrightarrow{p}^{\prime}\vert}{8\pi m_{B_{c}}^{2}}\left\vert\mathcal{A}(B_{c}^-\to J/\psi(\eta_{c})L^{-})\right\vert^{2}\,, \nn \\[0.2cm]
        \mathcal{B}(B_{c}^{-}\to J/\psi(\eta_{c}) L^{-}) &= \frac{\Gamma(B_{c}^{-}\to J/\psi(\eta_{c}) L^{-})}{\Gamma_\mathrm{tot}(B_c)}\,,
\end{align}
where $\Gamma_\mathrm{tot}(B_c)=1/\tau_{B_c}$ denotes the total decay width of the $B_{c}$ meson, and $\vert\overrightarrow{p}^{\prime}\vert$ is the magnitude of the three-momentum of the $J/\psi(\eta_{c})$ meson in the $B_c$-meson rest frame, with
\begin{align}
\vert\overrightarrow{p}^{\prime}\vert = \frac{\sqrt{\left[m_{B_{c}}^{2}-(m_{J/\psi(\eta_{c})}+m_{L})^{2}\right]\left[m_{B_{c}}^{2}-(m_{J/\psi(\eta_{c})}-m_{L})^{2}\right]}}{2m_{B_{c}}}\,.
\end{align}
Both the decay rates and branching fractions of $B_{c}^-\to J/\psi(\eta_{c})L^{-}$ decays are currently still plagued by large uncertainties from the $B_{c}^-\to J/\psi(\eta_{c})$ transition form factors~\cite{Harrison:2020gvo,Biswas:2023bqz}. There also exists a persistent tension between inclusive and exclusive determinations of the CKM matrix element $|V_{cb}|$~\cite{FlavourLatticeAveragingGroupFLAG:2024oxs,Bordone:2021oof}. To minimize the uncertainties brought by these input parameters, we can construct the ratios of the nonleptonic decay rates with respect to the corresponding differential semileptonic decay rates evaluated at $q^2=m_L^2$,
\begin{equation} \label{eq:nonlep2semilep}
R_{J/\psi(\eta_{c}) L} \equiv \frac{\Gamma(B_{c}^-\to J/\psi(\eta_{c}) L^-)}{d\Gamma(B_{c}^-\to J/\psi(\eta_{c}) \ell^-\bar{\nu}_{\ell})/dq^2\mid_{q^2=m_L^2}} = 6\pi^2\,|V_{uq}|^2\,f_L^2\,|a_1(J/\psi(\eta_{c}) L^-)|^2\, X_L\,,
\end{equation}
which, by construction, are free of the uncertainty related to $|V_{cb}|$. Neglecting the masses of light leptons $\ell=e,\,\mu$, we have exactly $X_L=1$ for a vector meson $L^-$~\cite{Neubert:1997uc,Beneke:2000ry}. For a light pseudoscalar meson $L^-$, on the other hand, $X_L$ depend on both the form-factor ratios and the kinematic factors; see eq.~(68) in ref.~\cite{Neubert:1997uc} for their explicit expressions. As demonstrated already in refs.~\cite{Huber:2016xod,Cai:2021mlt}, these ratios can be predicted more precisely and will be more suitable for probing NP beyond the SM. To compare with the LHCb measurement~\cite{LHCb:2014rck} of the ratio $R_{\pi/\mu\nu_{\mu}}^\mathrm{exp}$ given by eq.~\eqref{eq:LHCb-Rpimunu}, we will also consider the theoretical prediction of the same ratio: 
\begin{align}
R_{\pi/\mu\nu_{\mu}} = \frac{\mathcal{B}(B_{c}^{-}\to J/\psi \pi^{-})}{\mathcal{B}(B_{c}^{-}\to J/\psi \mu^{-}\bar{\nu}_{\mu})}\,,
\end{align}
where the semileptonic branching ratio integrated over the whole $q^2$ range is given as
\begin{equation}
   \mathcal{B}(B_{c}^{-}\to J/\psi \mu^{-}\bar{\nu}_{\mu}) = \int_{m_{\mu}^{2}}^{(m_{B_{c}}-m_{J/\psi})^2}\frac{d\mathcal{B}(B_{c}^{-}\to J/\psi \mu^{-}\bar{\nu}_{\mu})}{dq^2}dq^2\,.
\end{equation}
Throughout this paper, we will assume that the semileptonic $B_{c}^-\to J/\psi(\eta_{c}) \ell^-\bar{\nu}_{\ell}$ decays do not receive any NP contributions beyond the SM.

\section{Numerical results and discussions}
\label{sec:Numerical analysis}

In this section, we will update the SM predictions of the branching fractions of $B_{c}^-\to J/\psi(\eta_{c})L^{-}$ decays as well as the ratios $R_{J/\psi(\eta_{c}) L}$ and $R_{\pi/\mu\nu_{\mu}}$, by taking into account the latest input parameters and the NNLO vertex corrections to the hard-scattering kernels within the QCDF framework. The NP effects on these observables are then analyzed in a model-independent setup, by considering both the low- and high-scale scenarios. 

\subsection{Input parameters}

The observables of $B_{c}^-\to J/\psi(\eta_{c})L^{-}$ decays predicted within the QCDF framework depend on several input parameters, such as the strong coupling constant $\alpha_s$, the quark masses, the CKM matrix elements, the $B_c$-meson lifetime, the $B_{c}\to J/\psi(\eta_c)$ transition form factors, as well as the decay constants and the first two Gegenbauer moments of light mesons. Their values used throughout this paper together with the relevant references are collected in table~\ref{tab:inputs}. Especially, we will adopt the lattice QCD results of the $B_{c}\to J/\psi$ transition form factors~\cite{Harrison:2025kxm,Harrison:2020gvo,Harrison:2020nrv} extrapolated to the full kinematically allowed range with the Boyd-Grinstein-Lebed (BGL) parametrization~\cite{Boyd:1994tt,Boyd:1995sq,Boyd:1997kz}, while the values of the $B_{c}\to \eta_{c}$ transition form factors are obtained by using the heavy-quark spin symmetry relations between the associated form factors of $B_{c}\to J/\psi$ and $B_{c}\to \eta_{c}$ transitions, after parametrizing and extracting the possible symmetry breaking corrections~\cite{Biswas:2023bqz,Cohen:2019zev} (see also refs.~\cite{Colangelo:2022lpy,Jenkins:1992nb}). We have also used the three-loop running of $\alpha_s$, and the two-loop relation between pole and $\msbar$ mass to convert the top-quark pole mass $m_t^{\text{pole}}$ to the scale-invariant mass $\overline{m}_t(\overline{m}_t)$~\cite{Chetyrkin:2000yt}. To obtain the theoretical uncertainty of an observable, we vary each input parameter within its $1\sigma$ range and then add each individual uncertainty in quadrature. In addition, we have included the uncertainty due to the renormalization scale $\mu_b$ by varying it within the range $m_b/2\leq \mu_b \leq 2m_b$. 

%%%%%%%%%%%%%%%%%%%%%%%%%%%%%%%%%%%%%%%%%%%%%%%%%%%%%%%%%%%%%%%%%%%
\begin{table}[t]
\begin{center}	
\let\oldarraystretch=\arraystretch
\renewcommand*{\arraystretch}{1.33}
{\tabcolsep=0.781cm \begin{tabular}{|cccc|c|}
\hline\hline
\multicolumn{4}{|l|}{\textbf{\hspace{-0.18cm}QCD and electroweak parameters~~~\cite{ParticleDataGroup:2024cfk}}}
\\
\hline
  $G_F\,[10^{-5}\gev^{-2}]$
& $\alpha_s(m_Z)$
& $m_Z\,[\gev]$
& $m_W\,[\gev]$
\\
  $1.1663788$
& $0.1180 \pm 0.0009$
& $91.1880$
& $80.3692$
\\
\hline
\end{tabular}}

{\tabcolsep=0.263cm \begin{tabular}{|ccccc|}
\multicolumn{5}{|l|}{\hspace{0.5cm} \textbf{\hspace{-0.28cm}Quark masses\,[GeV]~~~\cite{ParticleDataGroup:2024cfk,ATLAS:2014wva}}}
\\
\hline
  $m_t^{\rm pole}$
& $\overline{m}_b(\overline{m}_b)$
& $\overline{m}_c(\overline{m}_c)$
& $\overline{m}_s(2\,\rm GeV)$
& $2\overline{m}_s/(\overline{m}_u+\overline{m}_d)$
\\
  $172.57 \pm 0.29$
& $4.183\pm 0.007$
& $1.2730 \pm 0.0046$
& $0.0935\pm 0.0008$
& $27.33_{-0.14}^{+0.18}$
\\
\hline
\end{tabular}}
{\tabcolsep=1.211cm \begin{tabular}{|ccc|}
\multicolumn{3}{|l|}{\hspace{-0.7cm} \textbf{CKM matrix elements~~~\cite{Charles:2004jd,ckm2018}}}
\\
\hline
  $|V_{ud}|$
& $|V_{us}|$
& $|V_{cb}|[10^{-3}]$
\\
  $0.9744129_{-0.0000513}^{+0.0000096}$
& $0.224791_{-0.000098}^{+0.000170} $
& $42.41_{-1.51}^{+0.40}$
\\
\hline
\end{tabular}}
{\tabcolsep=0.688cm \begin{tabular}{|cccc|}
		\multicolumn{4}{|l|}{\textbf{\boldmath Lifetimes and masses of $B_{c}$, $J/\psi$ and $\eta_c$ mesons~~~\cite{ParticleDataGroup:2024cfk,HFLAV:2022esi}}}
		\\
		\hline
		$\tau_{B_{c}}\,[{\rm ps}]$
		& $m_{B_{c}}\,[\mev]$
		& $m_{J/\psi}\,[\mev]$
		& $m_{\eta_c}\,[\mev]$
		\\
		$0.510 \pm 0.009$
		& $6274.47\pm 0.32$
		& $3096.900\pm 0.006$
		& $2984.1\pm 0.4$
		\\
		\hline
\end{tabular}}
{\tabcolsep=0.760cm \begin{tabular}{|cccc|}
		\multicolumn{4}{|l|}{\hspace{-0.2cm} \textbf{\boldmath $B_{c}\to J/\psi$ transition form factors evaluated at $q^{2}=0$~~~\cite{Harrison:2025kxm}}}
		\\
		\hline
		$V$
		& $A_{0}$
		& $A_{1}$
		& $A_{2}$
		\\
		\hline
		$0.751 \pm 0.058$
		& $0.465\pm 0.025$
		& $0.455\pm 0.019$
		& $0.437\pm 0.071$
		\\
		\hline
\end{tabular}}
{\tabcolsep=0.953cm \begin{tabular}{|cccc|}
		\multicolumn{4}{|l|}{\hspace{-0.4cm} \textbf{\boldmath $B_{c}\to \eta_{c}$ transition form factors~~~\cite{Biswas:2023bqz}}}
		\\
		\hline
		$f_{0}(m_{\pi}^2)$
		& $f_{0}(m_{K}^2)$
		& $f_{+}(m_{\rho}^2)$
		& $f_{+}(m_{K^{*}}^2)$
		\\
		\hline
		$0.45 \pm 0.20$
		& $0.45\pm 0.20$
		& $0.47\pm 0.20$
		& $0.48\pm 0.20$
		\\
		\hline
\end{tabular}}
{\tabcolsep=0.21cm \begin{tabular}{|c|cccc|c|}
\multicolumn{6}{|l|}{\hspace{0.38cm} \textbf{Masses, decay constants and Gegenbauer moments of light mesons}}
\\
\hline
\multicolumn{1}{|c|}{}
& $\pi^-$
& $K^-$
& $\rho^-$
& $K^{\ast-}$
&
\\
\hline
\multicolumn{1}{|c|}{$m_{L}\,[\mev]$}
& $139.57$
& $493.68$
& $775.26$
& $891.67$
& \textbf{{\cite{ParticleDataGroup:2024cfk}}}
\\
\hline
\multicolumn{1}{|c|}{$f_L\,[\mev]$}
& $130.2 \pm 1.7$
& $155.6 \pm 0.4$
& $216 \pm 6$
& $211 \pm 7$
& \\
\multicolumn{1}{|c|}{$f_{L}^{\perp}\,[\mev]$}
& --
& --
& $160 \pm 11$
& $163 \pm 8$
&\textbf{\cite{Rosner:2015wva,Straub:2015ica,Dimou:2012un}}
\\
\hline
\multicolumn{1}{|c|}{$a_1^L$}
& --
& $-0.0525^{+0.0033}_{-0.0031}$
& --
& $0.06 \pm 0.04$
&
\\
\multicolumn{1}{|c|}{$a_2^L$}
& $0.116^{+0.019}_{-0.020}$
& $0.106^{+0.015}_{-0.016}$
& $0.17 \pm 0.07$
& $0.16 \pm 0.09$
& \textbf{\cite{Straub:2015ica,Dimou:2012un,Arthur:2010xf,Bali:2019dqc}}
\\
\hline\hline
\end{tabular}}
 \caption{Summary of the input parameters used throughout this paper. The Gegenbauer moments of light mesons are evaluated at the scale $\mu=2\,\gev$, and the values of the CKM matrix elements are taken from the CKMfitter determinations updated in 2018 with only tree-level inputs~\cite{ckm2018}. \label{tab:inputs}}
\end{center}
\end{table}
%%%%%%%%%%%%%%%%%%%%%%%%%%%%%%%%%%%%%%%%%%%%%%%%%%%%%%%%%%%%%%%%%%%

To update our previous analysis of the class-I $\bar B_{(s)}^0\to D_{(s)}^{(*)+} L^-$ decays made in ref.~\cite{Cai:2021mlt}, we have also included the latest Belle measurements of the branching fractions of $\bar{B}^0\to D^{(*)+} \pi(K)^{-}$ decays~\cite{Belle:2021udv,Belle:2022afp}, as well as the updated fitting results of the $B_{(s)}\to D_{(s)}^{(*)}$ transition form factors~\cite{Bordone:2019vic,Bernlochner:2022ywh,Cui:2023jiw}, with their $q^2$ dependence given by the BGL parametrization~\cite{Boyd:1994tt,Boyd:1995sq,Boyd:1997kz}. Specifically, we will adopt the combined ``$\text{Lattice} \oplus \text{LCSR}$'' fitting results of these transition form factors~\cite{Cui:2023jiw}, which are obtained by performing a simultaneous fitting to both the lattice QCD data points~\cite{FermilabLattice:2021cdg,Na:2015kha,MILC:2015uhg,Harrison:2021tol,McLean:2019qcx} and the updated LCSR computations~\cite{Wang:2017jow,Gao:2021sav}. To update the experimental values of the ratios $R_{(s)L}^{(*)}=\Gamma(\bar B_{(s)}^0\to D_{(s)}^{(*)+} L^-)/d\Gamma(\bar B_{(s)}^0\to D_{(s)}^{(*)+} \ell^- \bar{\nu}_{\ell})/dq^2\mid_{q^2=m_L^2}$ constructed in ref.~\cite{Cai:2021mlt}, we extract the differential decay rates of the semileptonic $B_{(s)}^0\to D_{(s)}^{(*)+} \ell^- \bar{\nu}_{\ell}$ decays evaluated at $q^2=m_L^2$, by taking as input the combined ``$\text{Lattice} \oplus \text{LCSR} \oplus \text{Exp.}$'' fitting results of the $B_{(s)}\to D_{(s)}^{(*)}$ transition form factors~\cite{Cui:2023jiw}, where the experimental data points from refs.~\cite{Belle:2018ezy,Belle:2015pkj,Aaij:2020hsi,LHCb:2020hpv} have been taken into account during the fit. 

\subsection{SM predictions of the observables of \texorpdfstring{$B_{c}^-\to J/\psi(\eta_{c})L^{-}$}{B{c} to J/psi(eta c)L} decays}

%%%%%%%%%%%%%%%%%%%%%%%%%%%%%%%%%%%%%%%%%%%%%%%%%%%%%%%%%%%%%%%%%%%%%
\begin{figure}[t]
	\centering
	\includegraphics[width=0.49\textwidth]{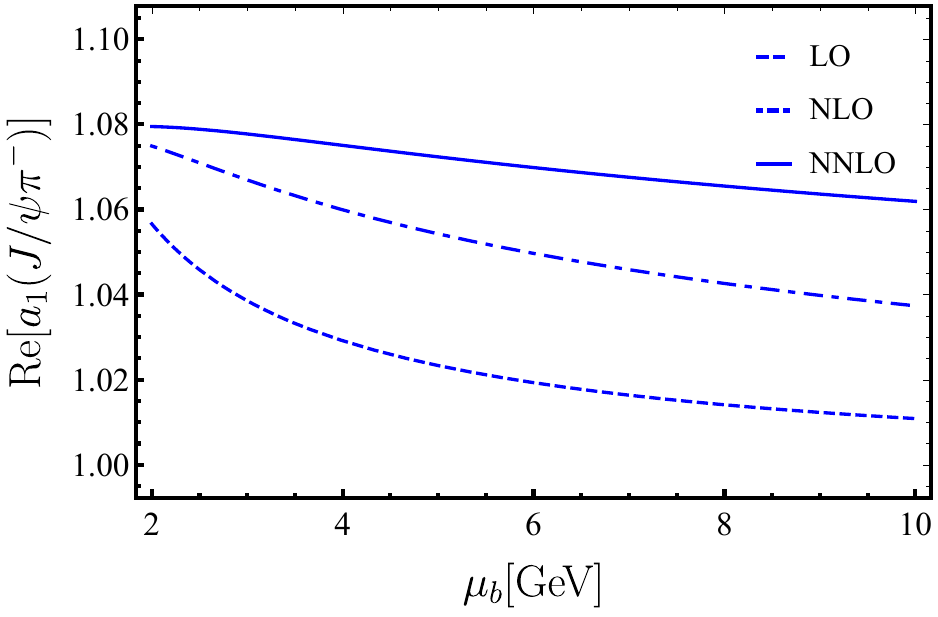}\;
        \includegraphics[width=0.49\textwidth]{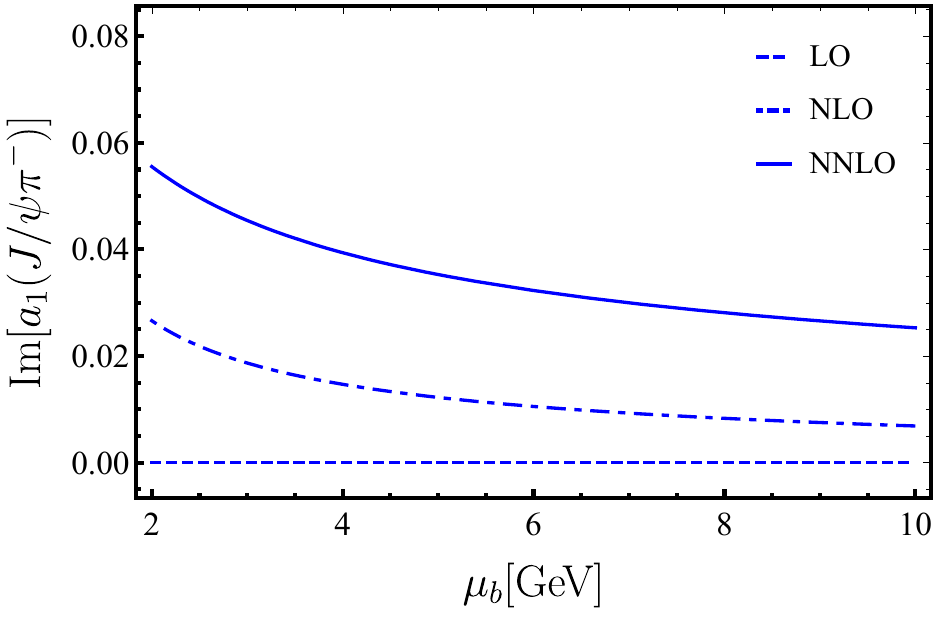}
	\caption{Dependence of the colour-allowed tree amplitude $a_{1}(J/\psi \pi^{-})$ on the renormalization scale $\mu_b$, where the dashed, dash-dotted and solid curves refer to the SM predictions at the LO, NLO and NNLO in $\alpha_s$, respectively. \label{scaledepa1} }
\end{figure}
%%%%%%%%%%%%%%%%%%%%%%%%%%%%%%%%%%%%%%%%%%%%%%%%%%%%%%%%%%%%%%%%%%%%%

\begin{figure}[t]
   \centering
   \includegraphics[width=0.55\textwidth]{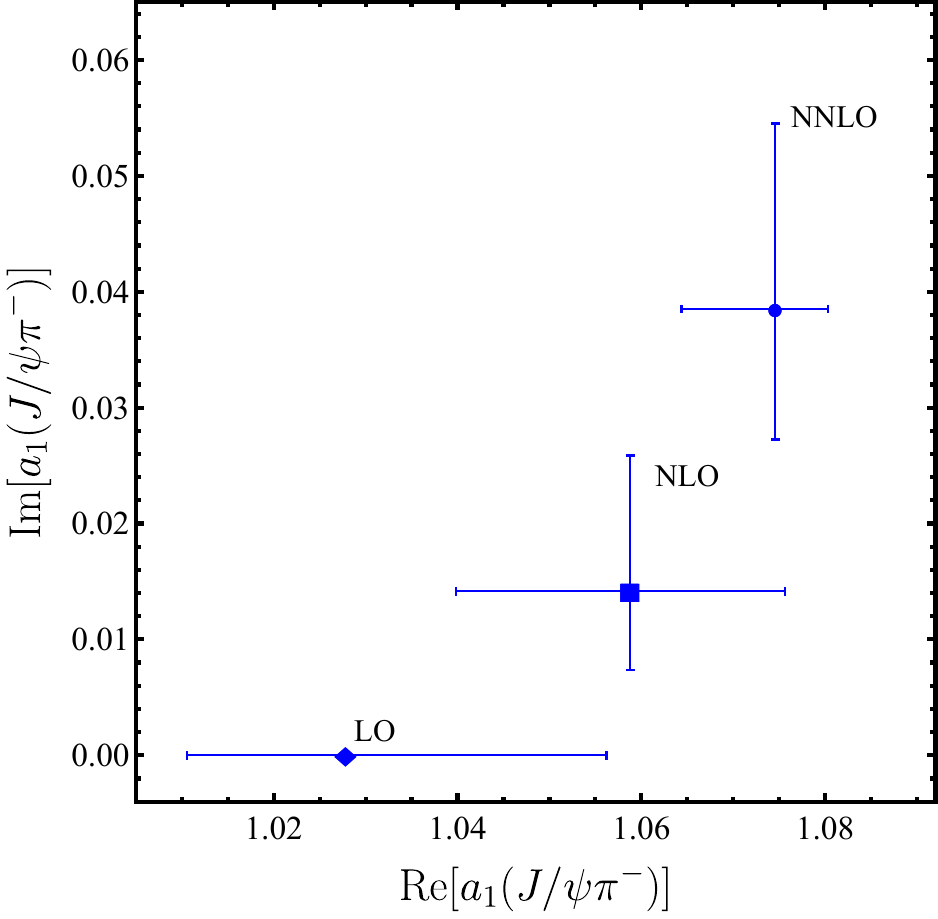}
   \caption{Graphical representation of $a_{1}(J/\psi \pi^{-})$ in the complex plane up to different orders in $\alpha_s$. The theoretical error estimates are also indicated.} \label{Fig:a1}
\end{figure}

It is known that, due to the truncation of the perturbative expansion, the colour-allowed tree amplitudes $a_1(J/\psi(\eta_{c})L^{-})$ defined in eq.~\eqref{eq:a1_effective} depend on the renormalization scale $\mu_b$, which can be, however, reduced by including the higher-order perturbative QCD corrections within the QCDF framework~\cite{Beneke:2009ek,Bell:2007tv,Bell:2009nk,Bell:2015koa,Huber:2016xod,Bell:2020qus}. Thus, before giving the updated SM predictions of the observables of $B_{c}^-\to J/\psi(\eta_{c})L^{-}$ decays, let us show in figure~\ref{scaledepa1} the scale dependence of $a_1(J/\psi\pi^{-})$ up tp different orders in $\alpha_s$. It can be clearly seen that a considerable stabilization of the scale dependence is achieved for the real part $\mathrm{Re}[a_1(J/\psi\pi^{-})]$, after taking into account the NNLO vertex correction. The reduction is, however, absent for the imaginary part $\mathrm{Im}[a_1(J/\psi\pi^{-})]$, because it is zero at tree level and the two-loop contribution is actually a NLO effect and, moreover, large. Our numerical result for $a_{1}(J/\psi \pi^{-})$ up to the NNLO in $\alpha_s$ is given by 
\begin{align} \label{eq:a1value}
    a_{1}(J/\psi \pi^{-}) &= 1.028 + [0.031+0.014i]_{\rm{NLO}} + [0.016+0.024i]_{\rm{NNLO}} \nonumber\\[0.2cm]
    &=(1.075_{-0.010}^{+0.006})+(0.039_{-0.011}^{+0.016})i\,,
\end{align}
where the number without bracket in the first line is the LO result, which has no imaginary part, while the following two numbers represent the NLO and NNLO corrections, respectively. The total errors in the second line comprise the uncertainties, added in quadrature, from the variation of the renormalization scale $\mu_b \in [m_b/2, 2 m_b]$, the quark masses, the Gegenbauer moments, and the QCD coupling $\alpha_s(m_Z)$, as collected in table~\ref{tab:inputs}. A graphical representation of eq.~\eqref{eq:a1value} in the complex plane up to different orders in $\alpha_s$ is also shown in figure~\ref{Fig:a1}. We can see that both the NLO and NNLO contributions add always constructively to the LO result, with the new two-loop correction amounting to approximately $61.5\%$ and $1.5\%$ of the total imaginary and real parts of $a_{1}(J/\psi \pi^{-})$. However, we must emphasize that the sizable NNLO correction to the imaginary part does not indicate a breakdown of the perturbative expansion, but is due to the fact that the imaginary part vanishes at LO, and its NLO term is colour suppressed and proportional to the small Wilson coefficient $\mC_{1}(\mu)$, while the colour suppression gets lifted and the large Wilson coefficient $\mC_{2}(\mu)$ re-enters at the NNLO. Furthermore, the impact from the imaginary part of $a_{1}(J/\psi \pi^{-})$ is only marginal for the branching ratios, as will be demonstrated later. It is also generally expected that no further dynamical mechanisms and hence large perturbative corrections could arise at even higher orders in $\alpha_s$~\cite{Beneke:2009ek,Huber:2016xod}.

%%%%%%%%%%%%%%%%%%%%%%%%%%%%%%%%%%%%%%%%%%%%%%%%%%%%%%%%%%%%%%%%%%%
%%%%%%%%%%%%%%%%%%%%%%%%%%%%%%%%%%%%%%%%%%%%%%%%%%%%%%%%%%%%%%%%%%%

With the relevant input parameters collected in table~\ref{tab:inputs}, our updated SM predictions of the branching ratios of $ B_{c}^-\to J/\psi(\eta_{c})L^{-}$ decays up to different orders in $\alpha_{s}$ are presented in table~\ref{tab:br}. To discuss the impacts from the higher-order perturbative corrections on the branching ratios, we define
\begin{align}
    \delta_{\text{NLO}} &= \frac{\mathcal{B}^\text{NLO} - \mathcal{B}^\text{LO}}{\mathcal{B}^\text{LO}}, \nonumber \\[0.15cm]
    \delta_{\text{NNLO}} &= \frac{\mathcal{B}^\text{NNLO} - \mathcal{B}^\text{LO}}{\mathcal{B}^\text{LO}},
\end{align}
where $\mathcal{B}^\text{LO}$ are the branching ratios predicted at the LO, while $\mathcal{B}^\text{NLO}$ and $\mathcal{B}^\text{NNLO}$ represent the same observables up to the NLO and NNLO in $\alpha_s$, respectively. We can see that, relative to the LO results, the branching ratios of these decays are always enhanced by the NLO and NNLO corrections, with a relative amount given by $\delta_{\text{NLO}} \approx +6\%$ and $\delta_{\text{NNLO}} \approx +9\%$, respectively. As discussed for $a_{1}(J/\psi \pi^{-})$, the sizable NNLO corrections to the branching ratios cannot be regarded as a breakdown of the perturbative expansion, and its dynamical reason is just because the NLO correction to the colour-allowed tree amplitudes $a_{1}(J/\psi (\eta_c) L^-)$ is colour-suppressed and proportional to the small Wilson coefficient $\mC_{1}(\mu_b)$, while at the NNLO the colour suppression gets lifted and the large Wilson coefficient $\mC_{2}(\mu_b)$ re-enters. A similar behaviour has also been observed in ref.~\cite{Huber:2016xod} for $\bar{B}_{(s)}^0\to D_{(s)}^{(*)+} L^-$ decays.

%%%%%%%%%%%%%%%%%%%%%%%%%%%%%%%%%%%%%%%%%%%%%%%%%%%%%%%%%%%%%%%%%%%
\begin{table}[t]
\begin{center}
\tabcolsep 0.23cm
\let\oldarraystretch=\arraystretch
\renewcommand*{\arraystretch}{1.5}	
\begin{tabular}{lcccccccccccc}
\hline \hline
%Decay mode  & $\mathrm{LO}$ & $\mathrm{NLO}$ & $\mathrm{NNLO}$ & \cite{AbdEl-Hady:1999jux} & \cite{Ebert:2003cn}&\cite{Sun:2007ei} & \cite{Chang:2014jca} &\cite{Issadykov:2018myx} &\cite{Cheng:2021svx}&\cite{Biswas:2023bqz}\\
Decay mode & $\mathrm{LO}$ & $\mathrm{NLO}$ & $\mathrm{NNLO}$ & \cite{Issadykov:2018myx} & \cite{Cheng:2021svx} & \cite{Biswas:2023bqz} \\
\hline
  $B_{c}^{-}\to J/\psi\pi^-$
& $0.741$
& $0.786_{-0.105}^{+0.094}$
& $0.811_{-0.105}^{+0.094}$
%& $1.09$
%& $0.600$
%& $1.19$
%& $1.11_{-0.10}^{+0.10}$
& $1.09_{-0.22}^{+0.22}$
& $1.47_{-0.07}^{+0.08}$
& $0.693_{-0.092}^{+0.092}$
\\
  $B_{c}^{-}\to J/\psi K^-$
& $0.562$
& $0.597_{-0.077}^{+0.068}$
& $0.618_{-0.077}^{+0.068}$
%& $0.810$
%& $0.468$
%& $--$
%& $0.854_{-0.073}^{+0.072}$  
& $0.81_{-0.18}^{+0.18}$
& $1.11_{-0.06}^{+0.06}$
& $0.520_{-0.070}^{+0.070}$
\\ 
$B_{c}^{-}\to J/\psi \rho^-$
& $2.093$
& $2.221_{-0.439}^{+0.436}$
& $2.291_{-0.447}^{+0.444}$
%& $3.12$
%& $1.61$
%& $--$
%& $3.21_{-0.52}^{+0.57}$
& $1.82_{-0.37}^{+0.37}$
& $8.29_{-0.56}^{+0.52}$
& $--$
\\ 
$B_{c}^{-}\to J/\psi K^{*-}$
& $1.072$
& $1.136_{-0.221}^{+0.219}$
& $1.167_{-0.225}^{+0.222}$
%& $1.76$
%& $0.981$
%& $--$
& $1.16_{-0.27}^{+0.27}$
& $4.67_{-0.36}^{+0.32}$ 
& $--$
%& $--$
\\
$B_{c}^{-}\to \eta_{c} \pi^-$
& $0.743$
& $0.788_{-0.549}^{+0.857}$
& $0.809_{-0.563}^{+0.880}$
%& $1.42$
%& $0.829$
%& $1.28$
%& $1.05_{-0.09}^{+0.09}$
& $2.03_{-0.41}^{+0.41}$
& $--$
& $0.63_{-0.40}^{+0.40}$
\\ 
$B_{c}^{-}\to \eta_{c} K^-$
& $0.558$
& $0.592_{-0.412}^{+0.644}$
& $0.611_{-0.425}^{+0.664}$
%& $0.468$
%& $1.07$
%& $--$
%& $0.823_{-0.070}^{+0.070}$ 
& $1.52_{-0.27}^{+0.27}$
& $--$
& $0.48_{-0.31}^{+0.31}$
\\ 
$B_{c}^{-}\to \eta_{c} \rho^-$
& $2.027$
& $2.151_{-1.456}^{+2.225}$
& $2.209_{-1.494}^{+2.285}$
%& $3.34$
%& $2.05$
%& $--$
%& $2.58_{-0.41}^{+0.45}$ 
& $2.81_{-0.56}^{+0.56}$
& $--$
& $1.80_{-1.10}^{+1.10}$
\\ 
$B_{c}^{-}\to \eta_{c} K^{*-}$
& $1.040$
& $1.102_{-0.736}^{+1.113}$
& $1.128_{-0.753}^{+1.139}$
%& $1.78$
%& $1.07$
%& $--$
%& $1.53_{-0.19}^{+0.20}$ 
& $1.69_{-0.36}^{+0.36}$
& $--$
& $0.95_{-0.58}^{+0.58}$
\\ 
\hline\hline
\end{tabular}
\caption{Updated SM predictions of the branching ratios (in units of $10^{-3}$ for $b\to c\bar{u}d$ and $10^{-4}$ for $b\to c\bar{u}s$ transitions) of $B_{c}^-\to J/\psi(\eta_{c})L^{-}$ decays up to the NNLO in $\alpha_s$. Recent results obtained in the CCQM~\cite{Issadykov:2018myx}, the LCSR~\cite{Cheng:2021svx}, and the NRQCD factorization~\cite{Biswas:2023bqz} are also given, as a comparison. \label{tab:br}}
\end{center}
\end{table}
%%%%%%%%%%%%%%%%%%%%%%%%%%%%%%%%%%%%%%%%%%%%%%%%%%%%%%%%%%%%%%%%%%%

%%%%%%%%%%%%%%%%%%%%%%%%%%%%%%%%%%%%%%%%%%%%%%%%%%%%%%%%%%%%%%%%%%%%%%
\begin{table}[t]
\tabcolsep 0.38cm
\let\oldarraystretch=\arraystretch
\renewcommand*{\arraystretch}{1.5}
\begin{center}
\begin{tabular}{lccccc}
\hline \hline
Ratio & ${\rm LO}$ & ${\rm NLO}$ & ${\rm NNLO}$& Exp. \\
\hline
  $R_{J/\psi\pi}$
& $\phantom{-}1.001$
& $\phantom{-}1.063_{-0.047}^{+0.044}$
& $\phantom{-}1.096_{-0.036}^{+0.032}$
& $--$
  \\ 
  $R_{J/\psi K}$
& $\phantom{-}0.713$
& $\phantom{-}0.757_{-0.029}^{+0.027}$
& $\phantom{-}0.784_{-0.019}^{+0.015}$
& $--$
 \\ 
  $R_{J/\psi\rho}$
& $\phantom{-}2.414$
& $\phantom{-}2.562_{-0.173}^{+0.172}$
& $\phantom{-}2.642_{-0.160}^{+0.159}$
& $--$
 \\
  $R_{J/\psi K^{*}}$
& $\phantom{-}1.178$
& $\phantom{-}1.248_{-0.095}^{+0.095}$
& $\phantom{-}1.283_{-0.090}^{+0.090}$
& $--$
 \\ 
  $R_{\eta_{c}\pi}$
& $\phantom{-}1.008$
& $\phantom{-}1.069_{-0.047}^{+0.046}$
& $\phantom{-}1.098_{-0.035}^{+0.032}$
& $--$
  \\ 
  $R_{\eta_{c} K}$
& $\phantom{-}0.774$
& $\phantom{-}0.822_{-0.067}^{+0.103}$
& $\phantom{-}0.848_{-0.064}^{+0.103}$
& $--$
 \\ 
  $R_{\eta_{c}\rho}$
& $\phantom{-}2.772$
& $\phantom{-}2.940_{-0.194}^{+0.193}$
& $\phantom{-}3.021_{-0.176}^{+0.175}$
& $--$
 \\
  $R_{\eta_{c} K^{*}}$
& $\phantom{-}1.408$
& $\phantom{-}1.492_{-0.111}^{+0.111}$
& $\phantom{-}1.527_{-0.103}^{+0.104}$
& $--$
 \\ 
$R_{\pi/\mu\nu_{\mu}}$
&$\phantom{-}0.0482$
&$\phantom{-}0.0511_{-0.0051}^{+0.0051}$
&$\phantom{-}0.0527_{-0.0051}^{+0.0050}$
&$\phantom{-}0.0469\pm0.0054$~\cite{LHCb:2014rck}
\\
\hline \hline
\end{tabular}
\caption{SM predictions and experimental values (if available) of the ratios $R_{J/\psi(\eta_{c}) L}$ (in units of $\rm GeV^{2}$ for $b\to c\bar{u}d$ and $10^{-1}~\rm GeV^{2}$ for $b\to c\bar{u}s$ transitions) and $R_{\pi/\mu\nu_{\mu}}$. \label{tab:nonlep2semilep}}
\end{center}
\end{table}
%%%%%%%%%%%%%%%%%%%%%%%%%%%%%%%%%%%%%%%%%%%%%%%%%%%%%%%%%%%%%%%%%%%%%%

For the branching ratios of $B_{c}^-\to J/\psi(\eta_{c})L^{-}$ decays, the resulting theoretical uncertainties are still dominated by that of the $B_{c}\to J/\psi$ and $B_{c}\to \eta_{c}$ transition form factors~\cite{Harrison:2025kxm,Harrison:2020gvo,Harrison:2020nrv,Biswas:2023bqz}. To see this clearly and as a comparison, we have also shown in table~\ref{tab:br} the recent results obtained in the covariant confined quark model (CCQM)~\cite{Issadykov:2018myx}, the LCSR~\cite{Cheng:2021svx}, and the NRQCD factorization approach~\cite{Biswas:2023bqz}, where the $B_{c}\to J/\psi$ and $B_{c}\to \eta_{c}$ transition form factors are calculated within the CCQM in ref.~\cite{Issadykov:2018myx} and in the LCSR approach in ref.~\cite{Cheng:2021svx} respectively, while these hadronic parameters adopted in ref.~\cite{Biswas:2023bqz} are exactly the same as what we are using here. The effective colour-allowed tree coefficients $a_{1}(J/\psi(\eta_{c})L^{-})$ in the former two references have further been adjusted to our NNLO results. In this way, one can see that the different inputs of these hadronic parameters have sensible impacts on the predicted branching ratios, and our results are relatively more close to the recent predictions obtained in ref.~\cite{Biswas:2023bqz}. Thus, accurate experimental measurements of these branching ratios are crucial to further test these different approaches for the form-factor calculations.

Finally, we present in table~\ref{tab:nonlep2semilep} our SM predictions of the ratios $R_{J/\psi(\eta_{c}) L}$ and $R_{\pi/\mu\nu_{\mu}}$ up to different orders in $\alpha_s$. It can be seen that, by construction, these ratios have relatively smaller theoretical uncertainties, due to the exact cancellation of the CKM matrix element $V_{cb}$ and a significant reduction of their dependence on the $B_{c}\to J/\psi$ and $B_{c}\to \eta_{c}$ transition form factors. Thus, these ratios provide a particularly clean and direct method to test the factorization hypothesis~\cite{Bjorken:1988kk,Neubert:1997uc,Beneke:2000ry,Huber:2016xod}, and they are also ideal for probing the different NP scenarios behind these class-I nonleptonic decays~\cite{Cai:2021mlt,Fleischer:2010ca}. It is particularly interesting to note that our SM predictions of the ratio $R_{\pi/\mu\nu_{\mu}}$ agree well with the LHCb measurement~\cite{LHCb:2014rck} within errors, which could be used to further narrow down the NP parameter space allowed by the ratios $R_{(s)L}^{(*)}$~\cite{Cai:2021mlt}, as will be detailed in the next subsection.

\subsection{Model-independent analysis}
\label{subsec:model-independent}

With our prescription for the effective weak Hamiltonian given by eq.~\eqref{eq:Hamiltonian}, possible NP effects would be signaled by the nonzero NP Wilson coefficients $C_{i}(\mu)$ that accompany the NP four-quark operators. As a model-independent analysis, we will use the currently available experiment data on the ratios $R_{\pi/\mu\nu_{\mu}}$ and $R_{(s)L}^{(*)}$ to constrain the NP Wilson coefficients $C_{i}(\mu)$, both at the characteristic scale $\mu_b=m_{b}$ (low-scale scenario) and at the electroweak scale $\mu_W=m_{W}$ (high-scale scenario). 

\subsubsection{Updated allowed ranges of \texorpdfstring{$C_i(m_b)$}{Ci(mb)} under the \texorpdfstring{$R_{(s)L}^{(\ast)}$}{R[(s)L]{(ast)}} constraints}

Firstly, we give in table~\ref{tab:constraint} the updated allowed ranges of the NP Wilson coefficients $C_i(m_b)$ under the individual and combined (last column) constraints from the ten ratios $R_{(s)L}^{(\ast)}$ varied within $1\sigma$ ($68.27\%$ confidence level (C.L.)) and $2\sigma$ ($95.45\%$ C.L.) error bars, respectively. For more details, we refer the readers to ref.~\cite{Cai:2021mlt}. It can be seen that, after considering the latest Belle measurements of the branching fractions of $\bar{B}^0\to D^{(*)+} \pi(K)^{-}$ decays~\cite{Belle:2021udv,Belle:2022afp} and the updated fitting results of the $B_{(s)}\to D_{(s)}^{(*)}$ transition form factors~\cite{Bordone:2019vic,Bernlochner:2022ywh,Cui:2023jiw}, the deviations observed between the SM predictions and the experimental measurements of the branching ratios of $\bar{B}_{(s)}^0\to D_{(s)}^{(*)+} L^-$ decays can be only explained by the model-independent NP four-quark operators with the $(1+\gamma_{5}) \otimes (1-\gamma_{5})$ and $(1+\gamma_{5}) \otimes (1+\gamma_{5})$ structures, while the solution with the $\gamma^\mu (1+\gamma_{5}) \otimes \gamma_\mu (1-\gamma_{5})$ structure does not work anymore, under the combined constraints from the ten ratios $R_{(s)L}^{(\ast)}$ at the $2\sigma$ level. This is mainly due to the most precise measurement of the branching fraction of $\bar{B}^0\to D^{*+} K^-$ decay by the Belle collaboration~\cite{Belle:2022afp}, $\mathcal{B}(\bar{B}^0\to D^{*+} K^-)=\left(2.22 \pm 0.06~(\text{stat}) \pm 0.08~(\text{syst})\right)\times 10^{-4}$, where both the statistical and systematic uncertainties have been significantly improved due to a larger dataset and better understanding of the detector. 

%%%%%%%%%%%%%%%%%%%%%%%%%%%%%%%%%%%%%%%%%%%%%%%%%%%%%%%%%%%%%%%%%%%%%%
\begin{landscape}
%\begin{sidewaystable}[htbp] 
\centering
	\tabcolsep 0.03cm
	\let\oldarraystretch=\arraystretch
        \renewcommand*{\arraystretch}{0.8}
	\fontsize{5.0}{12.0}\selectfont
	\LTcapwidth=\linewidth
	\begin{longtable}{|c|c|c|c|c|c|c|c|c|c|c|c|c|c|c|}
		\hline
		\diagbox[width=7.8em]{NP Coeff.}{C.L.$\backslash$Obs.}& C.L. &$R_{\pi}$&$R_{\pi}^{*}$ &$R_{\rho}$ &$R_{K}$ &$R_{K}^{*}$ &$R_{K^{*}}$ &$R_{s\pi}$&$R_{sK}$&$R_{s\pi}^{\ast}$&$R_{sK}^{\ast}$&Combined\cr
		\hline
		\multirow{2}{*}{$C_{1}^{VLL}$}
		&$1\sigma$ &$[-1.526,-1.102]$&$[-1.052,-0.652]$&$[-1.564,-0.365]$&$[-1.341,-0.986]
		$&$[-0.749,-0.378]$&$[-1.114,\phantom{+}0.299]$&$[-1.415,-0.778] $&$[-1.588,-1.033]$&$[-2.282,-0.640] $&$[-2.698,-1.121] $&$\varnothing$\cr
		\cline{2-13}
		&$2\sigma$ &$[-1.670,-0.974]$&$[-1.182,-0.531]$&$[-2.101,\phantom{+}0.025]$&$[-1.492,-0.855]
		$&$[-0.906,-0.236]$&$[-1.740,\phantom{+}0.751]$&$[-1.683,-0.556] $&$[-1.857,-0.818]$&$[-3.400,-0.044] $&$[-3.862,-0.550] $&$\varnothing$\cr
		\hline
		\multirow{2}{*}{$C_{2}^{VLL}$}
		&$1\sigma$ &$[-0.257,-0.190]$&$[-0.183,-0.115]$&$[-0.264,-0.064]$&$[-0.225,-0.171]
		$&$[-0.128,-0.065]$&$[-0.192,\phantom{+}0.054]$&$[-0.240,-0.135] $&$[-0.264,-0.178]$&$[-0.387,-0.113] $&$[-0.447,-0.195] $&$\varnothing$\cr
		\cline{2-13}
		&$2\sigma$ &$[-0.280,-0.169]$&$[-0.205,-0.094]$&$[-0.347,\phantom{+}0.004]$&$[-0.249,-0.148]
		$&$[-0.155,-0.040]$&$[-0.293,\phantom{+}0.136]$&$[-0.282,-0.097] $&$[-0.306,-0.142]$&$[-0.554,-0.008] $&$[-0.613,-0.096] $&$\varnothing$\cr
		\hline
		\multirow{2}{*}{$C_{1}^{VLR}$}
		&$1\sigma$ &$[\phantom{+}0.482,\phantom{+}0.653]$&$[\phantom{+}0.291,\phantom{+}0.464]$&$[-0.669,-0.162]$&$[\phantom{+}0.435,\phantom{+}0.572]
		$&$[\phantom{+}0.167,\phantom{+}0.326]$&$[-0.485,\phantom{+}0.135]$&$[\phantom{+}0.343,\phantom{+}0.607] $&$[\phantom{+}0.456,\phantom{+}0.673]$&$[\phantom{+}0.286,\phantom{+}0.982] $&$[\phantom{+}0.499,\phantom{+}1.143] $&$\varnothing$\cr
		\cline{2-13}
		&$2\sigma$ &$[\phantom{+}0.428,\phantom{+}0.711]$&$[\phantom{+}0.238,\phantom{+}0.520]$&$[-0.881,\phantom{+}0.011]$&$[\phantom{+}0.378,\phantom{+}0.634]
		$&$[\phantom{+}0.103,\phantom{+}0.394]$&$[-0.740,\phantom{+}0.341]$&$[\phantom{+}0.246,\phantom{+}0.716] $&$[\phantom{+}0.363,\phantom{+}0.780]$&$[\phantom{+}0.020,\phantom{+}1.409] $&$[\phantom{+}0.245,\phantom{+}1.571] $&$\varnothing$\cr
		\hline
		\multirow{2}{*}{$C_{2}^{VLR}$}
		&$1\sigma$ &$[\phantom{+}0.190,\phantom{+}0.257]$&$[\phantom{+}0.115,\phantom{+}0.183]$&$[-0.264,-0.064]$&$[\phantom{+}0.171,\phantom{+}0.225]
		$&$[\phantom{+}0.065,\phantom{+}0.128]$&$[-0.192,\phantom{+}0.054]$&$[\phantom{+}0.135,\phantom{+}0.240] $&$[\phantom{+}0.178,\phantom{+}0.264]$&$[\phantom{+}0.113,\phantom{+}0.387] $&$[\phantom{+}0.195,\phantom{+}0.447] $&$\varnothing$\cr
		\cline{2-13}
		&$2\sigma$ &$[\phantom{+}0.169,\phantom{+}0.280]$&$[\phantom{+}0.094,\phantom{+}0.205]$&$[-0.347,\phantom{+}0.004]$&$[\phantom{+}0.148,\phantom{+}0.249]
		$&$[\phantom{+}0.040,\phantom{+}0.155]$&$[-0.293,\phantom{+}0.136]$&$[\phantom{+}0.097,\phantom{+}0.282] $&$[\phantom{+}0.142,\phantom{+}0.306]$&$[\phantom{+}0.008,\phantom{+}0.554] $&$[\phantom{+}0.096,\phantom{+}0.613] $&$\varnothing$\cr
		\hline
		\multirow{2}{*}{$C_{1}^{VRR}$}
		&$1\sigma$ &$[\phantom{+}1.102,\phantom{+}1.526]$&$[-1.052,-0.652]$&$[-1.564,-0.365]$&$[\phantom{+}0.986,\phantom{+}1.341]
		$&$[-0.749,-0.378]$&$[-1.114,\phantom{+}0.299]$&$[\phantom{+}0.778,\phantom{+}1.415] $&$[\phantom{+}1.033,\phantom{+}1.588]$&$[-2.282,-0.640] $&$[-2.698,-1.121] $&$\varnothing$\cr
		\cline{2-13}
		&$2\sigma$ &$[\phantom{+}0.974,\phantom{+}1.670]$&$[-1.182,-0.531]$&$[-2.101,\phantom{+}0.025]$&$[\phantom{+}0.855,\phantom{+}1.492]
		$&$[-0.906,-0.236]$&$[-1.740,\phantom{+}0.751]$&$[\phantom{+}0.556,\phantom{+}1.683] $&$[\phantom{+}0.818,\phantom{+}1.857]$&$[-3.400,-0.044] $&$[-3.862,-0.550] $&$\varnothing$\cr
		\hline
		\multirow{2}{*}{$C_{2}^{VRR}$}
		&$1\sigma$ &$[\phantom{+}0.190,\phantom{+}0.257]$&$[-0.183,-0.115]$&$[-0.264,-0.064]$&$[\phantom{+}0.171,\phantom{+}0.225]
		$&$[-0.128,-0.065]$&$[-0.192,\phantom{+}0.054]$&$[\phantom{+}0.135,\phantom{+}0.240] $&$[\phantom{+}0.178,\phantom{+}0.264]$&$[-0.387,-0.113] $&$[-0.447,-0.195] $&$\varnothing$\cr
		\cline{2-13}
		&$2\sigma$ &$[\phantom{+}0.169,\phantom{+}0.280]$&$[-0.205,-0.094]$&$[-0.347,\phantom{+}0.004]$&$[\phantom{+}0.148,\phantom{+}0.249]
		$&$[-0.155,-0.040]$&$[-0.293,\phantom{+}0.136]$&$[\phantom{+}0.097,\phantom{+}0.282] $&$[\phantom{+}0.142,\phantom{+}0.306]$&$[-0.554,-0.008] $&$[-0.613,-0.096] $&$\varnothing$\cr
		\hline
		\multirow{2}{*}{$C_{1}^{VRL}$}
		&$1\sigma$ &$[-0.653,-0.482]$&$[\phantom{+}0.291,\phantom{+}0.464]$&$[-0.669,-0.162]$&$[-0.572,-0.435]
		$&$[\phantom{+}0.167,\phantom{+}0.326]$&$[-0.485,\phantom{+}0.135]$&$[-0.607,-0.343] $&$[-0.673,-0.456]$&$[\phantom{+}0.286,\phantom{+}0.982] $&$[\phantom{+}0.499,\phantom{+}1.143] $&$\varnothing$\cr
		\cline{2-13}
		&$2\sigma$ &$[-0.711,-0.428]$&$[\phantom{+}0.238,\phantom{+}0.520]$&$[-0.881,\phantom{+}0.011]$&$[-0.634,-0.378]
		$&$[\phantom{+}0.103,\phantom{+}0.394]$&$[-0.740,\phantom{+}0.341]$&$[-0.716,-0.246] $&$[-0.780,-0.363]$&$[\phantom{+}0.020,\phantom{+}1.409] $&$[\phantom{+}0.245,\phantom{+}1.571] $&$\varnothing$\cr
		\hline
		\multirow{2}{*}{$C_{2}^{VRL}$}
		&$1\sigma$ &$[-0.257,-0.190]$&$[\phantom{+}0.115,\phantom{+}0.183]$&$[-0.264,-0.064]$&$[-0.225,-0.171]
		$&$[\phantom{+}0.065,\phantom{+}0.128]$&$[-0.192,\phantom{+}0.054]$&$[-0.240,-0.135] $&$[-0.264,-0.178]$&$[\phantom{+}0.113,\phantom{+}0.387] $&$[\phantom{+}0.195,\phantom{+}0.447] $&$\varnothing$\cr
		\cline{2-13}
		&$2\sigma$ &$[-0.280,-0.169]$&$[\phantom{+}0.094,\phantom{+}0.205]$&$[-0.347,\phantom{+}0.004]$&$[-0.249,-0.148]
		$&$[\phantom{+}0.040,\phantom{+}0.155]$&$[-0.293,\phantom{+}0.136]$&$[-0.282,-0.097] $&$[-0.306,-0.142]$&$[\phantom{+}0.008,\phantom{+}0.554] $&$[\phantom{+}0.096,\phantom{+}0.613] $&$\varnothing$\cr
		\hline
		\multirow{2}{*}{$C_{1}^{SLL}$}
		&$1\sigma$ &$[\phantom{+}0.553,\phantom{+}0.749]$&$[-0.849,-0.533]$&$\varnothing$&$[\phantom{+}0.561,\phantom{+}0.741]$&$[-0.675,-0.348]$&$R $&$[\phantom{+}0.393,\phantom{+}0.697] $&$[\phantom{+}0.587,\phantom{+}0.871]$&$[-1.797,-0.523] $&$[-2.354,-1.025] $&$\varnothing$\cr
		\cline{2-13}
		&$2\sigma$ &$[\phantom{+}0.490,\phantom{+}0.815]$&$[-0.951,-0.435]$&$R $&$[\phantom{+}0.487,\phantom{+}0.820]$&$[-0.816,-0.217]$&$R $&$[\phantom{+}0.283,\phantom{+}0.822] $&$[\phantom{+}0.467,\phantom{+}1.009]$&$[-2.577,-0.036] $&$[-3.228,-0.505] $&$\varnothing$\cr
		\hline
		\multirow{2}{*}{$C_{2}^{SLL}$}
		&$1\sigma$ &$[\phantom{+}0.184,\phantom{+}0.250]$&$[-0.283,-0.178]$&$\varnothing$&$[\phantom{+}0.187,\phantom{+}0.247]$&$[-0.225,-0.116]$&$R $&$[\phantom{+}0.131,\phantom{+}0.232] $&$[\phantom{+}0.196,\phantom{+}0.290]$&$[-0.599,-0.174] $&$[-0.784,-0.342] $&$\varnothing$\cr
		\cline{2-13}
		&$2\sigma$ &$[\phantom{+}0.163,\phantom{+}0.272]$&$[-0.317,-0.145]$&$R $&$[\phantom{+}0.163,\phantom{+}0.273]$&$[-0.272,-0.072]$&$R $&$[\phantom{+}0.094,\phantom{+}0.274] $&$[\phantom{+}0.156,\phantom{+}0.336]$&$[-0.859,-0.012] $&$[-1.076,-0.168] $&$\varnothing$\cr
		\hline
		\multirow{2}{*}{$C_{1}^{SLR}$}
		&$1\sigma$ &$[-0.874,-0.644]$&$[\phantom{+}0.621,\phantom{+}0.990]$&$\varnothing$&$[-0.865,-0.653]$&$[\phantom{+}0.405,\phantom{+}0.788]$&$R $&$[-0.813,-0.459] $&$[-1.017,-0.684]$&$[\phantom{+}0.609,\phantom{+}2.096] $&$[\phantom{+}1.195,\phantom{+}2.746] $&$\varnothing$\cr
		\cline{2-13}
		&$2\sigma$ &$[-0.951,-0.571]$&$[\phantom{+}0.507,\phantom{+}1.110]$&$\leq0.366 $&$[-0.958,-0.568]$&$[\phantom{+}0.253,\phantom{+}0.952]$&$R $&$[-0.959,-0.329] $&$[-1.178,-0.544]$&$[\phantom{+}0.042,\phantom{+}3.006] $&$[\phantom{+}0.589,\phantom{+}3.765] $&$\varnothing$\cr
		\hline
		\multirow{2}{*}{$C_{2}^{SLR}$}
		&$1\sigma$ &$[-0.250,-0.184]$&$[\phantom{+}0.178,\phantom{+}0.283]$&$\varnothing$&$[-0.247,-0.187]$&$[\phantom{+}0.116,\phantom{+}0.225]$&$R $&$[-0.232,-0.131] $&$[-0.290,-0.196]$&$[\phantom{+}0.174,\phantom{+}0.599] $&$[\phantom{+}0.342,\phantom{+}0.784] $&$\varnothing$\cr
		\cline{2-13}
		&$2\sigma$ &$[-0.272,-0.163]$&$[\phantom{+}0.145,\phantom{+}0.317]$&$R $&$[-0.273,-0.163]$&$[\phantom{+}0.072,\phantom{+}0.272]$&$R $&$[-0.274,-0.094] $&$[-0.336,-0.156]$&$[\phantom{+}0.012,\phantom{+}0.859] $&$[\phantom{+}0.168,\phantom{+}1.076] $&$\varnothing$\cr
		\hline
		\multirow{2}{*}{$C_{1}^{SRR}$}
		&$1\sigma$ &$[-0.749,-0.553]$&$[-0.849,-0.533]$&$\varnothing$&$[-0.741,-0.561]$&$[-0.675,-0.348]$&$R $&$[-0.697,-0.393] $&$[-0.871,-0.587]$&$[-1.797,-0.523] $&$[-2.354,-1.025] $&$\varnothing$\cr
		\cline{2-13}
		&$2\sigma$ &$[-0.815,-0.490]$&$[-0.951,-0.435]$&$R $&$[-0.820,-0.487]$&$[-0.816,-0.217]$&$R $&$[-0.822,-0.283] $&$[-1.009,-0.467]$&$[-2.577,-0.036] $&$[-3.228,-0.505] $&$[-0.815,-0.505]$\cr
		\hline
		\multirow{2}{*}{$C_{2}^{SRR}$}
        &$1\sigma$&$[-0.250,-0.184]$&$[-0.283,-0.178]$&$\varnothing$&$[-0.247,-0.187]$&$[-0.225,-0.116]$&$R $&$[-0.232,-0.131] $&$[-0.290,-0.196]$&$[-0.599,-0.174] $&$[-0.784,-0.342] $&$\varnothing$\cr
		\cline{2-13}
		&$2\sigma$ &$[-0.272,-0.163]$&$[-0.317,-0.145]$&$R $&$[-0.273,-0.163]$&$[-0.272,-0.072]$&$R $&$[-0.274,-0.094] $&$[-0.336,-0.156]$&$[-0.859,-0.012] $&$[-1.076,-0.168] $&$[-0.272,-0.168]$\cr
		\hline
		\multirow{2}{*}{$C_{1}^{SRL}$}
		&$1\sigma$ &$[\phantom{+}0.644,\phantom{+}0.874]$&$[\phantom{+}0.621,\phantom{+}0.990]$&$\varnothing$&$[\phantom{+}0.653,\phantom{+}0.865]$&$[\phantom{+}0.405,\phantom{+}0.788]$&$R $&$[\phantom{+}0.459,\phantom{+}0.813] $&$[\phantom{+}0.684,\phantom{+}1.017]$&$[\phantom{+}0.609,\phantom{+}2.096] $&$[\phantom{+}1.195,\phantom{+}2.746] $&$\varnothing$\cr
		\cline{2-13}
		&$2\sigma$ &$[\phantom{+}0.571,\phantom{+}0.951]$&$[\phantom{+}0.507,\phantom{+}1.110]$&$\leq0.366 $&$[\phantom{+}0.568,\phantom{+}0.958]$&$[\phantom{+}0.253,\phantom{+}0.952]$&$R $&$[\phantom{+}0.329,\phantom{+}0.959] $&$[\phantom{+}0.544,\phantom{+}1.178]$&$[\phantom{+}0.042,\phantom{+}3.006] $&$[\phantom{+}0.589,\phantom{+}3.765] $&$\varnothing$\cr
		\hline
		\multirow{2}{*}{$C_{2}^{SRL}$}
		&$1\sigma$ &$[\phantom{+}0.184,\phantom{+}0.250]$&$[\phantom{+}0.178,\phantom{+}0.283]$&$\varnothing$&$[\phantom{+}0.187,\phantom{+}0.247]$&$[\phantom{+}0.116,\phantom{+}0.225]$&$R $&$[\phantom{+}0.131,\phantom{+}0.232] $&$[\phantom{+}0.196,\phantom{+}0.290]$&$[\phantom{+}0.174,\phantom{+}0.599] $&$[\phantom{+}0.342,\phantom{+}0.784] $&$\varnothing$\cr
		\cline{2-13}
		&$2\sigma$ &$[\phantom{+}0.163,\phantom{+}0.272]$&$[\phantom{+}0.145,\phantom{+}0.317]$&$R $&$[\phantom{+}0.163,\phantom{+}0.273]$&$[\phantom{+}0.072,\phantom{+}0.272]$&$R $&$[\phantom{+}0.094,\phantom{+}0.274] $&$[\phantom{+}0.156,\phantom{+}0.336]$&$[\phantom{+}0.012,\phantom{+}0.859] $&$[\phantom{+}0.168,\phantom{+}1.076] $&$[\phantom{+}0.168,\phantom{+}0.272]$\cr 
        \hline
        \caption{\fontsize{6.0}{10.0}\selectfont Updated allowed ranges of the NP Wilson coefficients $C_i(m_b)$ under the individual and combined constraints (last column) from the ratios $R_{(s)L}^{(\ast)}$ varied within $1\sigma$ and $2\sigma$ error bars, respectively. Here ``$\varnothing$'' represents an empty set and ``R'' the set of all real numbers within the plot ranges of $C_i(m_b)$. For more details, we refer the readers to ref.~\cite{Cai:2021mlt}. \label{tab:constraint} }
	\end{longtable}
%\end{sidewaystable}
\end{landscape}
%%%%%%%%%%%%%%%%%%%%%%%%%%%%%%%%%%%%%%%%%%%%%%%%%%%%%%%%%%%%%%%%%%%%%%%

\subsubsection{Low-scale scenario}

%%%%%%%%%%%%%%%%%%%%%%%%%%%%%%%%%%%%%%%%%%%%%%%%%%%%%%%%%%%%%%%%%%%%%
\begin{figure}[t]
	\centering
	\includegraphics[width=1.0\textwidth]{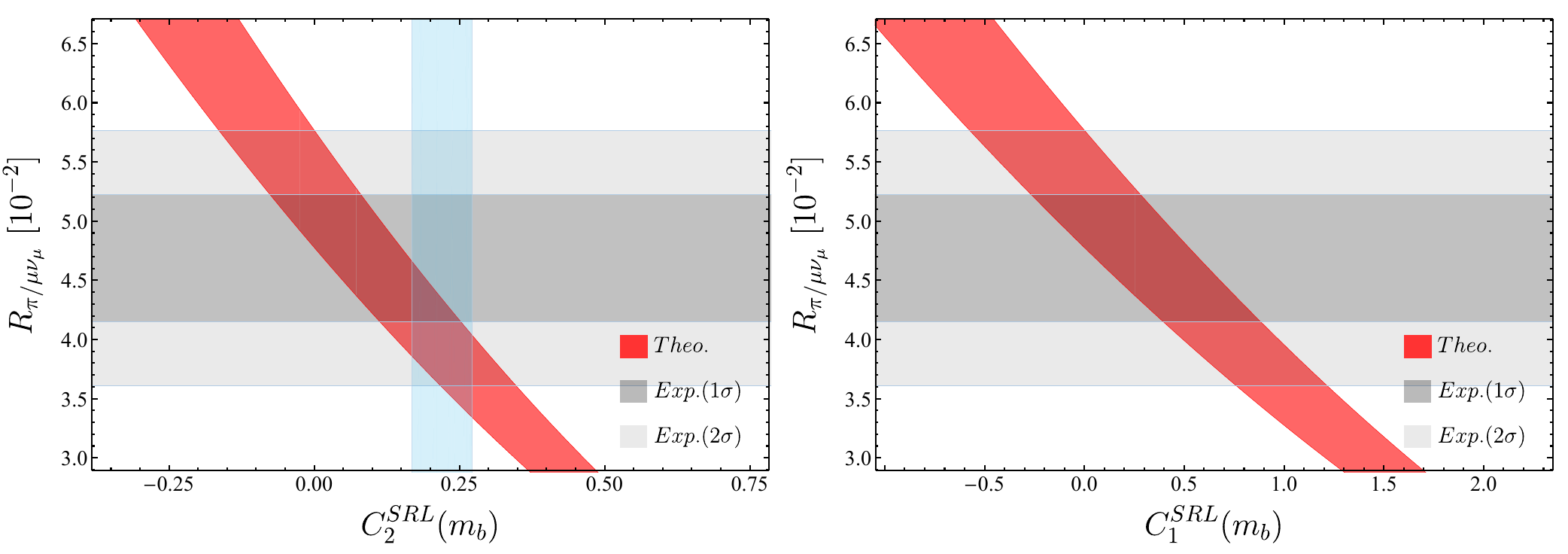}
        \caption{Constraints on the NP Wilson coefficients $C_{2}^{SRL}(m_b)$ (left) and $C_{1}^{SRL}(m_b)$ (right) from the ratios $R_{\pi/\mu\nu_{\mu}}$ (red) and $R_{(s)L}^{(*)}$ (cyan). The horizontal bounds represent the experimental ranges of the ratio $R_{\pi/\mu\nu_{\mu}}$ varied within $1\sigma$ (dark gray) and $2\sigma$ (light gray) error bars~\cite{LHCb:2014rck}. \label{MISRL} }
\end{figure}
%%%%%%%%%%%%%%%%%%%%%%%%%%%%%%%%%%%%%%%%%%%%%%%%%%%%%%%%%%%%%%%%%%%%%

%%%%%%%%%%%%%%%%%%%%%%%%%%%%%%%%%%%%%%%%%%%%%%%%%%%%%%%%%%%%%%%%%%%%%
\begin{figure}[t]
	\centering
	\includegraphics[width=1.0\textwidth]{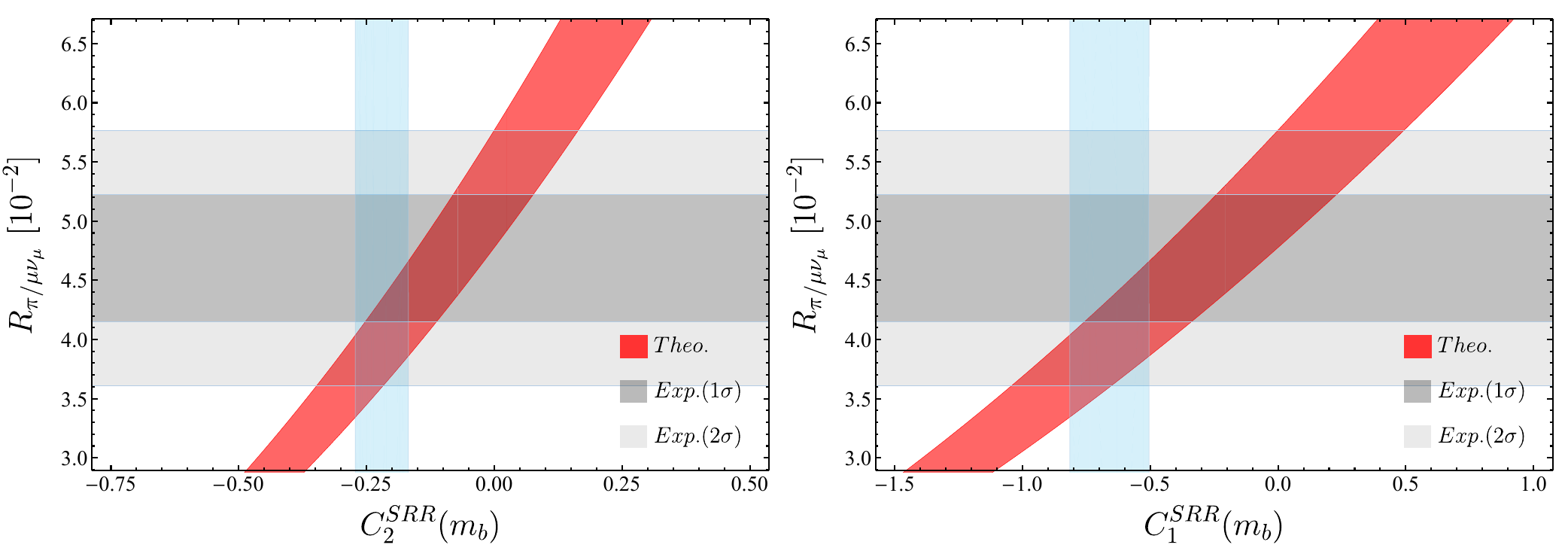}
	\caption{Same as in figure~\ref{MISRL} but for the NP Wilson coefficients $C_{2}^{SRR}(m_b)$ (left) and $C_{1}^{SRR}(m_b)$ (right). \label{MISRR} }
\end{figure}
%%%%%%%%%%%%%%%%%%%%%%%%%%%%%%%%%%%%%%%%%%%%%%%%%%%%%%%%%%%%%%%%%%%%%

To see if the ratio $R_{\pi/\mu\nu_{\mu}}$ can provide any complementary constraint on the NP Wilson coefficients $C_i(\mu)$ allowed by the ratios $R_{(s)L}^{(\ast)}$, we will firstly consider the low-scale scenario where only a single NP four-quark operator with one of the above Dirac structures is kept at the scale $\mu_b=m_{b}$. The resulting results are shown in figures~\ref{MISRL} and \ref{MISRR} for the NP Wilson coefficients $C_{1,2}^{SRL}(m_b)$ and $C_{1,2}^{SRR}(m_b)$, corresponding to the NP four-quark operators with the $(1+\gamma_{5})\otimes (1-\gamma_{5})$ and $(1+\gamma_{5})\otimes (1+\gamma_{5})$ structures, respectively. It can be seen that the allowed ranges of these Wilson coefficients by the ratios $R_{(s)L}^{(\ast)}$ can be marginally narrowed down by the ratio $R_{\pi/\mu\nu_{\mu}}$ varied within the $1\sigma$ error bar, with the final allowed ranges of $C_{2}^{SRL}(m_{b})$ and $C_{1,2}^{SRR}(m_{b})$ given, respectively, by
\begin{equation} \label{eq:NPWC_low}
\begin{aligned}
 & C_{2}^{SRL}(m_{b})\in [0.168,\,0.252]\,, \\[0.2cm]
 & C_{2}^{SRR}(m_{b})\in [-0.252,\,-0.168]\,, \qquad  & & C_{1}^{SRR}(m_{b})\in [-0.757,\,-0.505]\,,
\end{aligned}
\end{equation}
under the combined constraints from all these ratios. However, when varying the experimental data on the ratio $R_{\pi/\mu\nu_{\mu}}$ within the $2\sigma$ error bar, the resulting allowed ranges of $C_{2}^{SRL}(m_{b})$ and $C_{1,2}^{SRR}(m_{b})$ from $R_{\pi/\mu\nu_{\mu}}$ are much larger than that from the combined constraint from the ten ratios $R_{(s)L}^{(*)}$. This implies that more precise measurement of the ratio $R_{\pi/\mu\nu_{\mu}}$ is especially welcome from the LHC~\cite{Gouz:2002kk,Gao:2010zzc}.

\subsubsection{High-scale scenario}

%%%%%%%%%%%%%%%%%%%%%%%%%%%%%%%%%%%%%%%%%%%%%%%%%%%%%%%%%%%%%%%%%%%%%
\begin{figure}[t]
	\centering
	\includegraphics[width=1.0\textwidth]{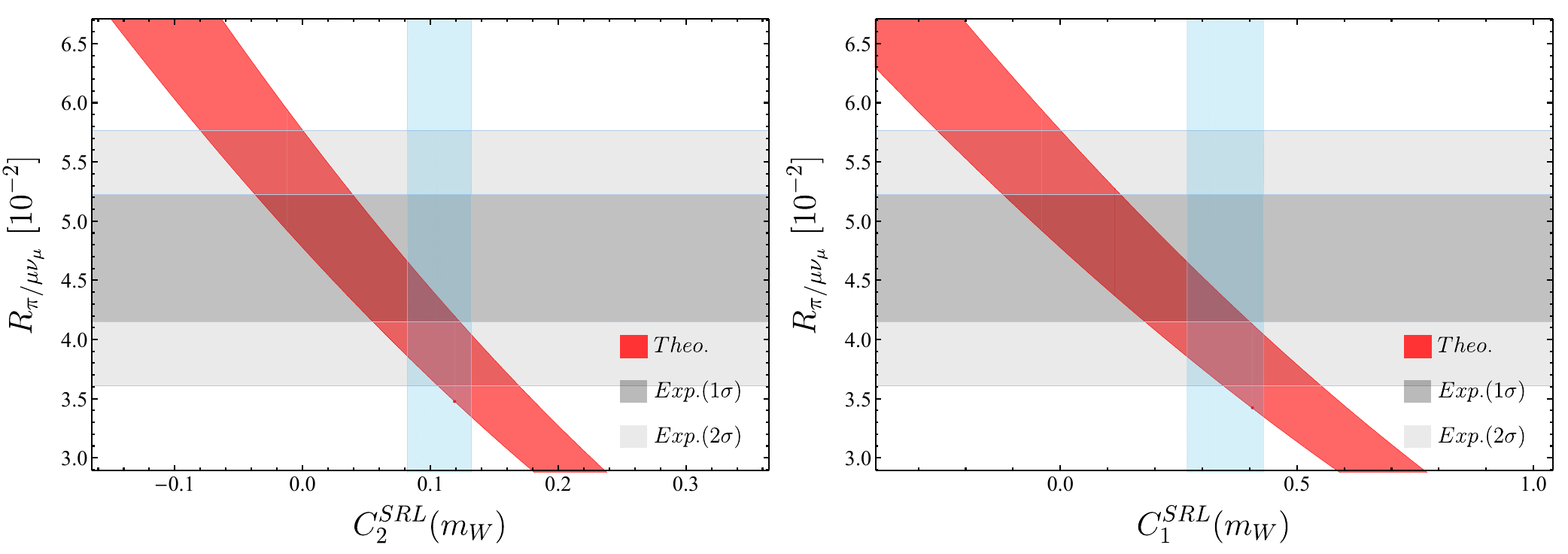}	
	\caption{Same as in figure~\ref{MISRL} but for the NP Wilson coefficients $C_{2}^{SRL}(m_W)$ (left) and $C_{1}^{SRL}(m_W)$ (right). \label{MISRLmW} }
\end{figure}
%%%%%%%%%%%%%%%%%%%%%%%%%%%%%%%%%%%%%%%%%%%%%%%%%%%%%%%%%%%%%%%%%%%%%

\begin{figure}[t]
	\centering
	\includegraphics[width=1.0\textwidth]{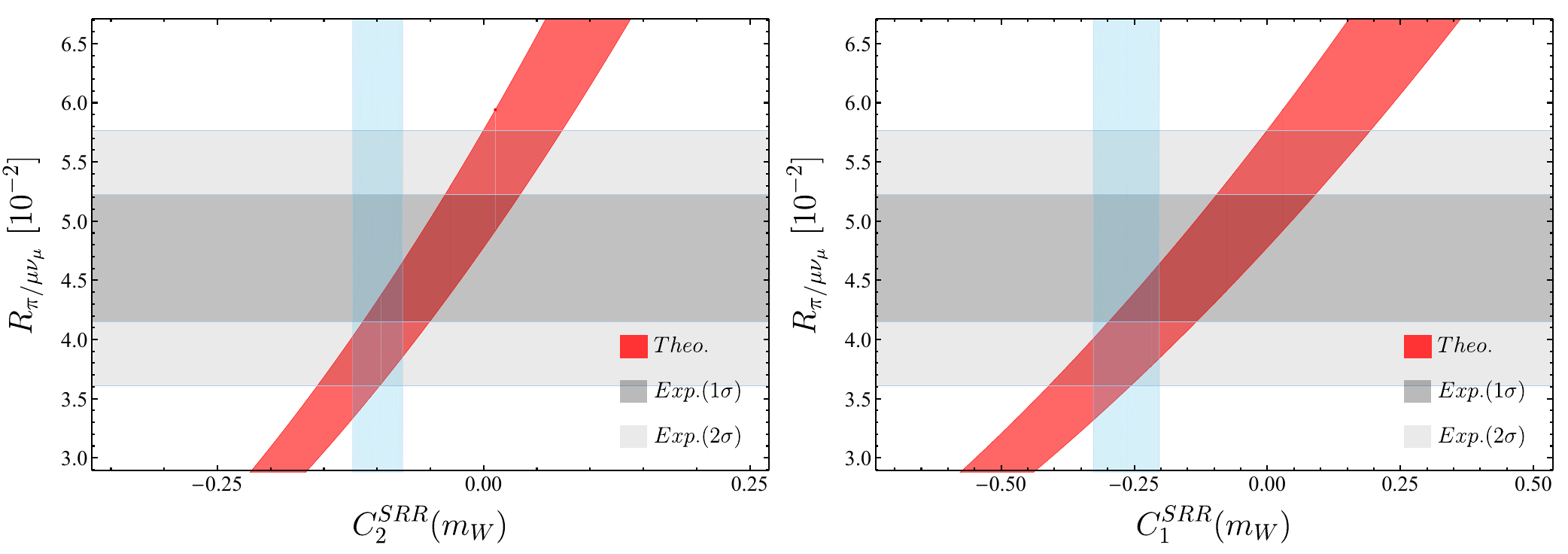}
	\caption{Same as in figure~\ref{MISRL} but for the NP Wilson coefficients $C_{2}^{SRR}(m_W)$ (left) and $C_{1}^{SRR}(m_W)$ (right). \label{MISRRmW} }
\end{figure}

We now turn to discuss the high-scale scenario, where the NP Wilson coefficients $C_i(\mu)$ are evaluated at the electroweak scale $\mu_W=m_{W}$. This is welcome for constructing specific NP models and correlating the low-energy constraints with the direct searches performed at high-energy frontiers. To this end, we must take into account the RG evolution of these short-distance Wilson coefficients from the electroweak scale $\mu_W$ down to the low-energy scale $\mu_b=m_{b}$. The most generic formulae for the RG equations satisfied by these NP Wilson coefficients $C_i(\mu)$ can be written as
\begin{equation} \label{eq:RGE}
    \mu\frac{dC_{j}(\mu)}{d\mu}=\gamma_{ij}(\mu)C_{i}(\mu)\,,
\end{equation}
where $\gamma_{ij}$ are the QCD ADMs of the corresponding NP four-quark operators, and their one- and two-loop expressions could be found, e.g., in ref.~\cite{Buras:2000if}. Solving eq.~\eqref{eq:RGE}, one can then obtain the evolution matrices $\hat{U}(\mu_b,\mu_W)$ connecting these NP Wilson coefficients at different scales, with
\begin{equation}
    \vec{C}(\mu_b)=\hat{U}(\mu_b,\mu_W)\,\vec{C}(\mu_W)\,.
\end{equation}
In this way, we can obtain the resulting constraints on the NP Wilson coefficients $C_{i}(m_{W})$ from the ratios $R_{\pi/\mu\nu_{\mu}}$ and $R_{(s)L}^{(*)}$, which are shown in figures~\ref{MISRLmW} and \ref{MISRRmW} for $C_{1,2}^{SRL}(m_{W})$ and $C_{1,2}^{SRR}(m_{W})$, respectively. We can see that the constraints from $R_{\pi/\mu\nu_{\mu}}$ and $R_{(s)L}^{(*)}$ are still complementary to each other. Numerically, the allowed ranges of the NP Wilson coefficients $C_{1,2}^{SRL}(m_{W})$ and $C_{1,2}^{SRR}(m_{W})$ under the combined constraints from $R_{\pi/\mu\nu_{\mu}}$ at the $1\sigma$ and $R_{(s)L}^{(*)}$ at the $2\sigma$ level are given, respectively, as
\begin{equation} \label{eq:NPWC_high}
\begin{aligned}
 & C_{2}^{SRL}(m_{W})\in [0.082,0.123]\,, \qquad & & C_{1}^{SRL}(m_{W})\in [0.268,0.400]\,, \\[0.2cm]
 & C_{2}^{SRR}(m_{W})\in [-0.114,-0.076]\,, \qquad & & C_{1}^{SRR}(m_{W})\in [-0.298,-0.203]\,.
\end{aligned}
\end{equation}
It is interesting to note that, due to large RG evolution effect, there exists now an allowed range of the NP Wilson coefficient $C_{1}^{SRL}(m_{W})$, although it is already ruled out at the scale $m_{b}$ by the combined constraints from the ten ratios $R_{(s)L}^{(*)}$ varied within the $2\sigma$ error bar. 

%%%%%%%%%%%%%%%%%%%%%%%%%%%%%%%%%%%%%%%%%%%%%%%%%%%%%%%%%%%%%%%%%%%%%
%%%%%%%%%%%%%%%%%%%%%%%%%%%%%%%%%%%%%%%%%%%%%%%%%%%%%%%%%%%%%%%%%%%%%

%%%%%%%%%%%%%%%%%%%%%%%%%%%%%%%%%%%%%%%%%%%%%%%%%%%%%%%%%%%%%%%%%%%%%%
\begin{table}[t]
\tabcolsep 0.45cm
\let\oldarraystretch=\arraystretch
\renewcommand*{\arraystretch}{1.5}
\begin{center}
\begin{tabular}{lcc}
\hline \hline
$C_{i}(m_{b})$ & Allowed ranges & Scenarios \\
\hline
  $C_{2}^{SRL}(m_{b})$
& $\phantom{+}0.210\pm 0.042$
& S1
  \\ 
  $C_{2}^{SRR}(m_{b})$
& $-0.210\pm 0.042$
& S2
  \\ 
  $C_{1}^{SRR}(m_{b})$
& $-0.631\pm 0.126$
& S3
 \\ 
\hline \hline
\end{tabular}
\caption{Allowed ranges of the NP Wilson coefficients $C_{i}(\mu)$ at the scale $m_{b}$ under the combined constraints from the ratios $R_{(s)L}^{(*)}$ and $R_{\pi/\mu\nu_{\mu}}$ varied within the $2\sigma$ and $1\sigma$ error bars, respectively. 
\label{tab:NP scenarios}}
\end{center}
\end{table}
%%%%%%%%%%%%%%%%%%%%%%%%%%%%%%%%%%%%%%%%%%%%%%%%%%%%%%%%%%%%%%%%%%%%%%

\subsection{Predictions of the observables in different NP scenarios}
\label{subsec:Predictions in NP scenarios}

Focusing on the low-energy scenario with the resulting allowed ranges of the NP Wilson coefficients $C_{2}^{SRL}(m_{b})$ and $C_{1,2}^{SRR}(m_{b})$ given by eq.~\eqref{eq:NPWC_low}, we now investigate how the three solutions affect the branching fractions of $B_{c}^-\to J/\psi(\eta_c) L^{-}$ decays and the ratios $R_{J/\psi(\eta_{c}) L}$. To this end, we assume again that only a single NP four-quark operator with one of the $(1+\gamma_{5}) \otimes (1-\gamma_{5})$ and $(1+\gamma_{5}) \otimes (1+\gamma_{5})$ structures is present in eq.~\eqref{eq:Hamiltonian}, with the central value and error bar of the corresponding Wilson coefficient given by the midpoint and half the length of the interval in eq.~\eqref{eq:NPWC_low}. We mark the three solutions as three different scenarios, and tabulated them in table~\ref{tab:NP scenarios}. The resulting branching fractions of $B_{c}^-\to J/\psi(\eta_c) L^{-}$ decays and the ratios $R_{J/\psi(\eta_{c}) L}$ are finally presented in table~\ref{tab:br1}. 

%%%%%%%%%%%%%%%%%%%%%%%%%%%%%%%%%%%%%%%%%%%%%%%%%%%%%%%%%%%%%%%%%%%%%%%%%%%%%%%%%%%%%%
\begin{table}[t]
\begin{center}
\tabcolsep 0.36cm
\let\oldarraystretch=\arraystretch
\renewcommand*{\arraystretch}{1.5}	
\begin{tabular}{lccccccccccc}
\hline \hline
Observables  & S1 & S2 &S3 &SM \\
\hline
  $\mathcal{B} (B_{c}^{-}\to J/\psi\pi^-)$
& $0.6188_{-0.1041}^{+0.0962}$
& $0.6188_{-0.1041}^{+0.0962}$
& $0.6186_{-0.1041}^{+0.0962}$
& $0.811_{-0.105}^{+0.094}$
\\
  $\mathcal{B} (B_{c}^{-}\to J/\psi K^-)$
& $0.4881_{-0.0769}^{+0.0706}$
& $0.4881_{-0.0769}^{+0.0706}$
& $0.4879_{-0.0769}^{+0.0706}$
& $0.618_{-0.077}^{+0.068}$
\\ 
  $\mathcal{B} (B_{c}^{-}\to J/\psi \rho^-)$
& $2.2906_{-0.4473}^{+0.4439}$
& $2.2906_{-0.4473}^{+0.4439}$
& $2.3126_{-0.4494}^{+0.4458}$
& $2.291_{-0.447}^{+0.444}$
\\ 
  $\mathcal{B} (B_{c}^{-}\to J/\psi K^{*-})$
& $1.1675_{-0.2248}^{+0.2220}$
& $1.1675_{-0.2248}^{+0.2220}$
& $1.1811_{-0.2261}^{+0.2232}$
& $1.167_{-0.225}^{+0.222}$
\\ 
  $\mathcal{B} (B_{c}^{-}\to \eta_{c}\pi^-)$
& $0.5158_{-0.3699}^{+0.5680}$
& $0.5158_{-0.3699}^{+0.5680}$
& $0.5154_{-0.3696}^{+0.5675}$
& $0.809_{-0.563}^{+0.880}$
\\
  $\mathcal{B} (B_{c}^{-}\to \eta_{c} K^-)$
& $0.4129_{-0.2939}^{+0.4532}$
& $0.4129_{-0.2939}^{+0.4532}$
& $0.4126_{-0.2937}^{+0.4529}$
& $0.611_{-0.425}^{+0.664}$
\\ 
  $\mathcal{B} (B_{c}^{-}\to \eta_{c} \rho^-)$
& $2.2093_{-1.4943}^{+2.2846}$
& $2.2093_{-1.4943}^{+2.2846}$
& $2.2032_{-1.4902}^{+2.2783}$
& $2.209_{-1.494}^{+2.285}$
\\ 
  $\mathcal{B} (B_{c}^{-}\to \eta_{c} K^{*-})$
& $1.1282_{-0.7526}^{+1.1390}$
& $1.1282_{-0.7526}^{+1.1390}$
& $1.1248_{-0.7503}^{+1.1356}$
& $1.128_{-0.753}^{+1.139}$
\\ 
\hline
  $R_{J/\psi\pi}$
& $0.8364_{-0.0932}^{+0.0902}$
& $0.8364_{-0.0932}^{+0.0902}$
& $0.8360_{-0.0933}^{+0.0902}$
& $1.096_{-0.036}^{+0.032}$
  \\ 
  $R_{J/\psi K}$
& $0.6190_{-0.0611}^{+0.0596}$
& $0.6190_{-0.0611}^{+0.0596}$
& $0.6188_{-0.0611}^{+0.0597}$
& $0.784_{-0.019}^{+0.015}$
 \\ 
  $R_{J/\psi\rho}$
& $2.6418_{-0.1601}^{+0.1588}$
& $2.6418_{-0.1601}^{+0.1588}$
& $2.6672_{-0.1597}^{+0.1589}$
& $2.642_{-0.160}^{+0.159}$
 \\
  $R_{J/\psi K^{*}}$
& $1.2830_{-0.0897}^{+0.0902}$
& $1.2830_{-0.0897}^{+0.0902}$
& $1.2979_{-0.0897}^{+0.0905}$
& $1.283_{-0.090}^{+0.090}$
 \\ 
  $R_{\eta_{c}\pi}$
& $0.7001_{-0.1232}^{+0.1240}$
& $0.7001_{-0.1232}^{+0.1240}$
& $0.6996_{-0.1233}^{+0.1241}$
& $1.098_{-0.035}^{+0.032}$
  \\ 
  $R_{\eta_{c} K}$
& $0.5731_{-0.0974}^{+0.1124}$
& $0.5731_{-0.0974}^{+0.1124}$
& $0.5728_{-0.0974}^{+0.1124}$
& $0.848_{-0.064}^{+0.103}$
 \\ 
  $R_{\eta_{c}\rho}$
& $3.0205_{-0.1764}^{+0.1747}$
& $3.0205_{-0.1764}^{+0.1747}$
& $3.0122_{-0.1760}^{+0.1744}$
& $3.021_{-0.176}^{+0.175}$
 \\
  $R_{\eta_{c} K^{*}}$
& $1.5273_{-0.1028}^{+0.1035}$
& $1.5273_{-0.1028}^{+0.1035}$
& $1.5226_{-0.1026}^{+0.1034}$
& $1.527_{-0.103}^{+0.104}$
 \\ 
\hline \hline
\end{tabular}
\caption{Predicted branching fractions (in units of $10^{-3}$ for $b\to c\bar{u}d$ and $10^{-4}$ for $b\to c\bar{u}s$ transitions) of $B_{c}^-\to J/\psi(\eta_{c})~L^{-}$ decays and the ratios $R_{J/\psi(\eta_{c}) L}$ (in units of $\rm GeV^{2}$ for $b\to c\bar{u}d$ and $10^{-1}~\rm GeV^{2}$ for $b\to c\bar{u}s$ transitions), in the S1, S2 and S3 scenarios defined in table~\ref{tab:NP scenarios}. The last column labeled with ``SM'' represents our NNLO predictions within the SM given in tables~\ref{tab:br} and \ref{tab:nonlep2semilep}. \label{tab:br1}}
\end{center}
\end{table}

It can be seen that some of the branching fractions and the ratios $R_{J/\psi(\eta_{c}) L}$ predicted in these three different NP scenarios can differ from the corresponding SM results, but are still plagued by large theoretical uncertainties. More precise data on $B_{c}^-\to J/\psi(\eta_{c})~L^{-}$ decays, as expected from the LHC with its high collision energy and high luminosity~\cite{Gouz:2002kk,Gao:2010zzc}, is therefore needed to further discriminate these different NP scenarios. It is also observed that the S1 and S2 scenarios give the same impacts on the observables of these decays and, for some processes like $B_{c}^-\to J/\psi(\eta_c) \rho^{-}$ and $B_{c}^-\to J/\psi(\eta_c) K^{*-}$, the SM predictions are even not changed by these two scenarios up to the NLO accuracy in QCD. This can be inferred from the analytical expressions of the hadronic matrix elements of the corresponding NP four-quark operators evaluated at the LO and NLO in $\alpha_s$, for different final states in $B_{c}^-\to J/\psi(\eta_c) L^{-}$ decays. Especially, within the QCDF approach and once the three-particle contributions to the light-meson LCDAs are neglected~\cite{Beneke:2003zv,Beneke:2006hg}, the hadronic matrix elements of the NP four-quark operators $\mQ_{2}^{SRL(R)} =\bigl[\bar{c}_{\alpha}(1+\gam_5)b_{\alpha}\bigr] \bigl[\bar{q}_{\beta}(1 \mp \gam_5)u_{\beta}\bigr]$ in S1~(S2) scenario, $\langle J/\psi(\eta_c) \rho(K^*)^{-}|\mQ_{2}^{SRL(R)}| B_{c}^-\rangle$, are zero up to the NLO in $\alpha_s$, while that of $\mQ_{1}^{SRR} =\bigl[\bar{c}_{\alpha}(1+\gam_5)b_{\beta}\bigr] \bigl[\bar{q}_{\beta}(1+\gam_5)u_{\alpha}\bigr]$ in S3 scenario, $\langle J/\psi(\eta_c) \rho(K^*)^{-}|\mQ_{1}^{SRR}| B_{c}^-\rangle$, is nonzero at the NLO in $\alpha_s$, for a longitudinally polarized vector meson $\rho(K^*)^{-}$. For further details, we refer the readers to refs.~\cite{Cai:2021mlt,Meiser:2024zea}. 

\section{Conclusions}
\label{sec:conclusions}

In this paper, motivated by the recent observation that the measured branching ratios of $\bar{B}_s^0\to D_s^+ \pi^-$ and $\bar{B}_d^0\to D^+ K^-$ decays deviate significantly from the latest SM predictions based on the QCDF approach, we have revisited the two-body hadronic $B_{c}^-\to J/\psi(\eta_{c})L^{-}$ decays, with $L=\pi, K^{(*)}, \rho$, within the same theoretical framework. Since these processes are mediated by the quark-level $b\to c \bar{u} d(s)$ transitions and hence dominated by the colour-allowed tree topology, the QCD factorization is generally expected to hold in the heavy-quark limit. We have extended the previous calculations~\cite{Sun:2007ei,Qiao:2012hp} and updated the SM predictions of the branching ratios of $B_{c}^-\to J/\psi(\eta_{c})L^{-}$ decays, by including the nonfactorizable vertex corrections to the hadronic matrix elements of the SM four-quark operators up to the NNLO in $\alpha_s$. It is found that, relative to the LO results, the branching ratios of these processes up to the NLO and NNLO corrections are always enhanced, with a relative amount given by $\delta_{\text{NLO}} = (\mathcal{B}^\text{NLO}-\mathcal{B}^\text{LO})/\mathcal{B}^\text{LO} \approx +6\%$ and $\delta_{\text{NNLO}} = (\mathcal{B}^\text{NNLO}-\mathcal{B}^\text{LO})/\mathcal{B}^\text{LO} \approx +9\%$, respectively. To minimize the uncertainties brought by the CKM matrix element $V_{cb}$ as well as the $B_{c}\to J/\psi$ and $B_{c}\to \eta_c$ transition form factors, we have also constructed the ratios $R_{J/\psi(\eta_{c}) L}$ and $R_{\pi/\mu\nu_{\mu}}$, which provide not only a particularly clean and direct method to test the factorization hypothesis, but also an ideal way for probing the different NP scenarios behind these class-I nonleptonic decays. 

Starting from the most general effective weak Hamiltonian given by eq.~\eqref{eq:Hamiltonian}, we have then investigated the NP effects on the observables of $B_{c}^-\to J/\psi(\eta_{c})L^{-}$ decays, by including the NLO vertex corrections to the hadronic matrix elements of the model-independent NP four-quark operators within the QCDF approach. We have also updated our previous analysis in ref.~\cite{Cai:2021mlt} by including the latest Belle measurements of the branching fractions of $\bar{B}^0\to D^{(*)+} \pi(K)^{-}$ decays and the updated fitting results of the $B_{(s)}\to D_{(s)}^{(*)}$ transition form factors. It is found that the deviations observed between the SM predictions and the experimental measurements of the branching ratios of $\bar{B}_{(s)}^0\to D_{(s)}^{(*)+} L^-$ decays can still be explained by the model-independent NP four-quark operators with the $(1+\gamma_{5}) \otimes (1-\gamma_{5})$ and $(1+\gamma_{5}) \otimes (1+\gamma_{5})$ structures, while the solution with the $\gamma^\mu (1+\gamma_{5}) \otimes \gamma_\mu (1-\gamma_{5})$ structure does not work anymore, under the combined constraints from the updated data on the ratios $R_{(s)L}^{(\ast)}$ at the $2\sigma$ level. Finally, to see if the LHCb measurement of the ratio $R_{\pi/\mu\nu_{\mu}}$ can provide any complementary constraint on the NP Wilson coefficients $C_i(\mu)$ allowed by $R_{(s)L}^{(\ast)}$, we have studied the variations of $R_{\pi/\mu\nu_{\mu}}$ with respect to the NP Wilson coefficients $C_{1,2}^{SRL}(m_{b})$ and $C_{1,2}^{SRR}(m_{b})$, assuming again that only a single NP four-quark operator with one of the above Dirac structures is present in eq.~\eqref{eq:Hamiltonian}. It is worth noting that the ratio $R_{\pi/\mu\nu_{\mu}}$, once measured precisely, can further narrow down the parameter space allowed by the ratios $R_{(s)L}^{(*)}$, and some of the predicted branching fractions of $B_{c}^-\to J/\psi(\eta_c) L^{-}$ decays and the ratios $R_{J/\psi(\eta_{c}) L}$ in the different allowed NP scenarios can differ from the corresponding SM values, but are still plagued by large theoretical uncertainties.

As the LHCb experiment is expected to produce around $5\times 10^{10}$ $B_c$-meson events per year~\cite{Gouz:2002kk,Gao:2010zzc}, we hope to obtain more precise measurements of the branching fractions of $B_{c}^-\to J/\psi(\eta_c) L^{-}$ decays as well as the ratios $R_{J/\psi(\eta_{c}) L}$ and $R_{\pi/\mu\nu_{\mu}}$, which can be exploited to further discriminate the different NP scenarios responsible for the  deviations observed in the branching ratios of $\bar{B}_s^0\to D_s^+ \pi^-$ and $\bar{B}_d^0\to D^+ K^-$ decays.

\section*{Acknowledgements}
We thank the authors of ref.~\cite{Cui:2023jiw} for useful discussions about the $B_{(s)}\to D_{(s)}^{(*)}$ transition form factors. This work is supported by the National Natural Science Foundation of China under Grant Nos.~12475094, 12135006, 12075097, 12061141006 and 12347175, the China Postdoctoral Foundation under Grant No.~GZB20230195, as well as the Natural Science Foundation of Hunan Province under Grant No.~2024JJ6179 and of Henan Province under Grant No.~242300421685. This work is also supported by the Fundamental Research Funds for the Central Universities under Grant No.~CCNU24AI003 and the Science and Technology Innovation Leading Talent Support Program of Henan Province.

%\appendix

\bibliographystyle{JHEP}
\bibliography{reference}

\end{document}